\documentclass[aps,pra,reprint,onecolumn,notitlepage,showkeys,showpacs,amsfonts,
               amsmath]{revtex4-1}
\usepackage{amsthm}
\usepackage{enumerate}

\theoremstyle{definition}
\newtheorem{example}{Example}[section]
\newtheorem{remark}{Remark}[section]

\newcommand{\Eqref}[1]{Equation~\eqref{#1}}

\newcommand{\secref}[1]{Section~\ref{#1}}

\newcommand{\Secref}[1]{Section~\ref{#1}}

\newcommand{\subsecref}[1]{Subsection~\ref{#1}}

\newcommand{\braket}[1]{\langle{#1}\rangle}

\newcommand{\lshad}{[\![}
\newcommand{\rshad}{]\!]}
\newcommand{\sdot}{\,\cdot\,}
\newcommand{\dd}{\mathrm{d}}
\newcommand{\vd}[1]{\dd{#1}\,}
\newcommand{\ud}{\,\dd}
\newcommand{\nbr}[1]{$#1$\nobreakdash-\hspace{0pt}}
\providecommand{\abs}[1]{\lvert#1\rvert}

\DeclareMathOperator{\tr}{tr}

\DeclareMathOperator{\id}{id}

\newcommand{\eqbreakadd}{}
\newcommand{\eqbreakmul}{}

\numberwithin{equation}{section}

\begin{document}

\title{Canonical quantization of classical mechanics in curvilinear
       coordinates.\\Invariant quantization procedure.}
\author{Maciej B{\l}aszak}
\email[Electronic address: ]{blaszakm@amu.edu.pl}
\author{Ziemowit Doma{\'n}ski}
\email[Electronic address: ]{ziemowit@amu.edu.pl}
\affiliation{Faculty of Physics, Adam Mickiewicz University\\
             Umultowska 85, 61-614 Pozna{\'n}, Poland}
\date{\today}

\begin{abstract}
In the paper is presented an invariant quantization procedure of classical
mechanics on the phase space over flat configuration space. Then, the passage to
an operator representation of quantum mechanics in a Hilbert space over
configuration space is derived. An explicit form of position and momentum
operators as well as their appropriate ordering in arbitrary curvilinear
coordinates is demonstrated. Finally, the extension of presented formalism onto
non-flat case and related ambiguities of the process of quantization are
discussed.
\end{abstract}



\maketitle

\section{Introduction}
\label{sec:1}
One of the fundamental problems in quantum mechanics is finding a consistent
quantization procedure of a classical system. In particular of associating to
classical observables operators defined on a Hilbert space. The usual way of
doing this is by using the knowledge of the form of a classical observable and
applying a Weyl quantization rule. More precisely, classical observables are
defined as real functions defined on a phase space. The Weyl quantization rule
states that to such function one associates an operator by formally replacing
$x$ and $p$ coordinates in classical observable with operators $\hat{q}$,
$\hat{p}$ of position and momentum, and symmetrically ordering them. By such
procedure one can quantize every classical Hamiltonian system. Note however,
that this procedure works only for systems whose phase space is
$\mathbb{R}^{2N}$. Moreover, quantization has to be performed in Cartesian
coordinates. Even in that well recognized case a natural question appears:
whether the Weyl quantization is a unique choice? In other words, whether there
are other quantization procedures which are consistent with physical
experiments.

The proper quantization procedure should be possible to perform in any
coordinate system. However, if we would take a classical system and naively
perform a quantization according to the Weyl quantization rule, for two
different canonical coordinates, then in general we would not get equivalent
quantum systems. Even more problems appear when quantization is performed in a
non-flat configuration space. As we will show later on this apparent
inconsistency of quantization in a flat case can be solved by a proper choice of
quantum observables in new coordinates, i.e. by performing an appropriate
deformation of classical observables written in new coordinates, or
alternatively by using different ordering rules of position and momentum
operators for different coordinates. We also discuss the admissible
quantizations in a Riemann space (non-flat configuration space) together with
an appropriate choice of quantum observables.

The problem with quantization in arbitrary coordinate system on a configuration
space was evident in early days of quantum mechanics. The majority of efforts
was related to invariant quantization of Hamiltonians quadratic in momenta.
The construction of a quantum Hamiltonian in flat and non-flat cases was
considered by many authors (see for example several relevant papers
\cite{Podolsky:1928,DeWitt:1952,DeWitt:1957,Gervais:1976,Carter:1977,%
Essen:1978,Dekker:1980,Liu:1992,Duval:2001}). Much less results concern
invariant quantization of Hamiltonians cubic in momenta
\cite{LoubonDjounga:2003,Duval:2005}. However, to our knowledge, there does not
exists general solution valid for any classical observable and canonical
transformation. In this paper we propose a consistent invariant quantization
procedure for a general flat case and admissible natural extensions of presented
procedure onto Riemann spaces.

The standard Hilbert space approach to quantum mechanics seems not very good for
quantizing classical Hamiltonian systems in different coordinates as it is hard
to keep control over the proper ordering of position and momentum operators.
More natural approach for this task appears to be a quantization over phase
space \cite{Moyal:1949,Bayen:1978a,Bayen:1978b}. This description of quantum
mechanics has mathematical structure similar to that of classical Liouville
mechanics (Hamiltonian mechanics in particular). This allows easy introduction
of concepts from classical theory to quantum counterpart, especially the concept
of canonical transformations of coordinates.

Therefore, in this paper we present the theory of quantization in general
canonical coordinates, related to Euclidean coordinates by nonlinear point
transformations. The basic concept is as follows. If we consider a classical
system described by a phase space $T^*E^N$ being a cotangent bundle to an
Euclidean space $E^N$, and take some Euclidean coordinate system $(x,p)$, then
to every classical observable $A(x,p)$ (function on the phase space) we can
associate an operator $\hat{A} = A_W(\hat{q},\hat{p})$ defined on the Hilbert
space $L^2(\mathbb{R}^N)$, being a symmetrically (Weyl) ordered function $A$ of
canonical operators of position $\hat{q}^i = x^i$ and momentum $\hat{p}_j =
-i\hbar\partial_{x^j}$. Such approach to quantization is equivalent with
approach based on deformation quantization where as the star-product we take the
Moyal product $\star_M^{(x,p)}$. So we can say that to Moyal product corresponds
Weyl ordering rule of canonical operators of position and momentum. If we now
choose some other canonical coordinate system $(x',p')$ we can write the Moyal
product $\star_M^{(x,p)}$ in this new coordinates receiving a new star-product
$\star^{(x',p')}$ (which in general will not be of the Moyal form), and new
operators of position and momentum $\hat{q}'^i = x'^i$, $\hat{p}'_j =
-i\hbar(\partial_{x'^j} + \frac{1}{2}\Gamma^k_{jk}(x'))$ (where $\Gamma^i_{jk}$
are Christoffel symbols of a standard linear connection on the configuration
space $E^N$), which construction we show explicitly. The key observation is that
to the new star-product $\star^{(x',p')}$ will correspond a new ordering rule.
In the paper we present a systematic way of constructing such new ordering.
In fact it is based on the observation that the star-product $\star^{(x',p')}$
is equivalent with the Moyal product $\star_M^{(x',p')}$ associated with the
coordinates $(x',p')$ by the corresponding morphism denoted by $S$. Thus, the
new ordering is equal to the Weyl ordering performed on an \nbr{S}deformed
classical function $A(x',p')$, i.e. $\hat{A}' = (S^{-1}A)_W(\hat{q}',\hat{p}')$.
The received quantization procedure in different coordinate systems is
consistent since, as we show, the operators $\hat{A}$ and $\hat{A}'$ are
unitarily equivalent.

Furthermore, we derive a covariant form of Hamiltonian operators in a case of
Hamiltonians quadratic and cubic in momenta and generalize the presented
formalism onto non-flat configuration spaces. Finally, we consider a family of
star-products which do not take the Moyal form in any coordinate system. Using
this family of star-products we present a discussion on the ambiguity of the
quantization process.

The paper in organized as follows. In \secref{sec:7} we review a classical
Liouville mechanics. In \secref{sec:2} we present a quantization of a classical
Hamiltonian system on a phase space in a coordinate independent way.
\Secref{sec:4} contains the passage to an operator representation in a Hilbert
space over configuration space, i.e. to a standard representation of quantum
mechanics in arbitrary curvilinear coordinates. \Secref{sec:5} contains remarks
how to use the presented formalism to construct admissible quantizations of
classical systems in Riemann spaces as well as remarks on dealing with some
ambiguity of the process of quantization.

\section{Classical Liouville mechanics}
\label{sec:7}
In this section the basics of the classical Liouville mechanics will be
reviewed. It is done using the language familiar to that used in quantum
mechanics, in order to make a quantization process more transparent.

\subsection{Hamiltonian systems}
\label{subsec:7.1}
A classical Hamiltonian system is a pair composed of a real Poisson manifold
$(M,\mathcal{P})$ ($\mathcal{P}$ being a Poisson tensor) and a smooth real
function $H$ defined on $M$. The function $H$ is a Hamilton function which
governs the time evolution. The Poisson manifold $(M,\mathcal{P})$ represents
a phase space, where dynamics takes place. A phase space can be naturally
induced from a configuration space $\mathcal{Q}$ as a cotangent bundle
$T^*\mathcal{Q}$ to $\mathcal{Q}$.

Using the Poisson tensor $\mathcal{P}$ a particular Lie bracket, called a
Poisson bracket, can be defined on a space $C^\infty(M,\mathbb{C})$ of all
(complex valued) smooth functions on $M$:
\begin{equation*}
\{f,g\}_{\mathcal{P}} := \mathcal{P}(\dd{f},\dd{g}), \quad f,g \in C^\infty(M).
\end{equation*}
The space $C^\infty(M)$ has also a structure of a commutative algebra with
involution, where multiplication $\cdot$ is a point-wise product of functions
and an involution is the complex conjugation. The double algebra with involution
$(C^\infty(M),\cdot,\{\sdot,\sdot\},\bar{\ })$ is called a Poisson algebra and
will be denoted by $\mathcal{A}_C$.

The geometric structure of the Poisson manifold $(M,\mathcal{P})$ is fully given
by a Poisson algebra $\mathcal{A}_C$. In particular two Poisson manifolds are
diffeomorphic iff the corresponding Poisson algebras are isomorphic.

Self-adjoint, with respect to the involution $f \mapsto \bar{f}$, elements from
$C^\infty(M)$, i.e. real valued functions, correspond to measurable quantities
and are called observables. They form a real subalgebra of the Poisson algebra
$\mathcal{A}_C$.

\subsection{Classical states}
\label{subsec:7.2}
In classical Liouville mechanics states are defined as probability distributions
defined on a phase space $M$, i.e. as `generalized' functions $\rho$ defined on
$M$ satisfying
\begin{enumerate}
\item  $\rho = \bar{\rho}$ (self-conjugation),
\item  $\displaystyle \int_M \rho \ud{\Omega} = 1$ (normalization),
\item  $\displaystyle \int_M \bar{f} \cdot f \cdot \rho \ud{\Omega} \ge 0$ for
$f \in C^\infty(M)$ $\iff$ $\rho \ge 0$ (positivity),
\end{enumerate}
where $\dd{\Omega}$ is a Liouville measure. States in the form of Dirac delta
distributions $\delta(\xi - \xi_0)$ play a distinguished role. We call them
pure states representing a situation when the localization of the system on the
phase space is known precisely. As a result the Liouville mechanics reduces to a
classical Hamiltonian mechanics. Observe that every pure state cannot be written
as a convex linear combination of some other states, i.e. there do not exist two
different states $\rho_1$ and $\rho_2$ such that $p\rho_1 + (1 - p)\rho_2$ is a
pure state for some $p \in (0,1)$. A converse statement is also true, namely, a
state which cannot be written as a convex linear combination of some other
states is a pure state. Note that pure states $\delta(\xi - \xi_0)$ can be
identified with points $\xi_0$ of the phase space.

In a case when the phase space $M$ is induced by a configuration space
$\mathcal{Q}$ of the form of a pseudo-Euclidean space $E^{r,s}$ with metric
signature $(r,s)$, i.e. $M = T^*E^{r,s}$, it is possible to characterize
classical states in a different way. In such special case it is possible to
introduce a multiplication between states, namely a convolution of functions
\begin{equation*}
f * g = \int_{\mathbb{R}^{2N}} f(\zeta)g(\xi - \zeta)
    \ud{\zeta}.
\end{equation*}
The space of states is closed with respect to such product. Pure states can be
defined then as those states which are idempotent
\begin{equation*}
\rho * \rho = \rho.
\label{eq:7.5}
\end{equation*}
Indeed, the idempotent states are precisely the Dirac delta distributions.
The general states can be described as convex linear combinations of pure
states
\begin{equation}
\rho = \sum_\lambda p_\lambda \rho^{(\lambda)}_{\text{pure}},
\label{eq:7.4}
\end{equation}
where $p_\lambda \ge 0$ and $\sum_\lambda p_\lambda = 1$. The summation in
\eqref{eq:7.4} can be in general integration performed over the phase space. In
such a case when $p \ge 0$ and $\int_M p \ud{\Omega} = 1$ we get that
\begin{equation*}
\rho(\xi) = \int_M p \rho^{(\xi_0)}_{\text{pure}} \ud{\Omega}
= \int_M p(\xi_0) \delta(\xi - \xi_0) \ud{\xi_0} = p(\xi),
\end{equation*}
and we reproduce the previous definition of states.

Quantities measured in experiment are expectation values of observables. For
a given observable $A \in C^\infty(M)$ and state $\rho$ the expectation value
of the observable $A$ in the state $\rho$ is defined by
\begin{equation*}
\braket{A}_\rho := \int_M A \cdot \rho \ud{\Omega}.
\end{equation*}
Note that an expectation value of the observable $A$ in a pure state
$\delta(\xi - \xi_0)$ is just equal $A(\xi_0)$.

\subsection{Time evolution of classical Hamiltonian systems}
\label{subsec:7.3}
For a given Hamiltonian system $(M,\mathcal{P},H)$ the Hamiltonian $H$ governs
the time evolution of the system. There are two dual points of view on the time
evolution. In the first one, called the classical Schr\"odinger picture, states
undergo the time development while observables do not. In the second one, called
the classical Heisenberg picture, states remain still whereas observables
undergo the time development. An equation of motion for states in the
Schr\"odinger picture takes the form
\begin{equation*}
\frac{\partial \rho}{\partial t}(t) - \{H,\rho(t)\} = 0
\end{equation*}
and is called a Liouville equation, whereas, a time evolution equation for
observables which do not explicitly depend on time, in the Heisenberg picture
reads
\begin{equation}
\frac{\dd{A}}{\dd{t}}(t) - \{A(t),H\} = 0.
\label{eq:7.1}
\end{equation}
Both presented approaches to the time development yield equal predictions
concerning the results of measurements, since
\begin{equation*}
\braket{A(0)}_{\rho(t)} = \braket{A(t)}_{\rho(0)}.
\end{equation*}

For pure state $\rho(\xi,t) = \delta(\xi - \xi_0(t))$, i.e. for classical
Hamiltonian mechanics:
\begin{equation*}
\frac{\partial \rho}{\partial t}(t) - \{H,\rho(t)\} = 0 \quad \Longrightarrow
\quad \frac{\dd{\xi_0^i}}{\dd{t}}(t) - \{\xi_0^i(t),H\} = 0.
\end{equation*}
Thus Schr\"odinger picture for pure states collapse onto a Heisenberg picture
for coordinates.

\subsection{Canonical coordinate system}
\label{subsec:7.4}
For every Poisson manifold there exists a distinguished class of local
coordinate systems called canonical (Darboux) coordinates. From the definition
these are the coordinates $(q^i,p_j)$ in which a Poisson tensor $\mathcal{P}$
takes the form
\begin{equation}
\mathcal{P} = \frac{\partial}{\partial q^i} \wedge\frac{\partial}{\partial p_i},
\qquad (\mathcal{P}^{ij}) = \begin{pmatrix}
0_N & \mathbb{I}_N \\
-\mathbb{I}_N & 0_N
\end{pmatrix}.
\label{eq:7.2}
\end{equation}

For a given canonical coordinate system $(q^i,p_j)$ functions
$Q^i(q,p) = q^i$,  $P_j(q,p) = p_j$
are observables of position and momentum associated with that system. Then the
condition \eqref{eq:7.2} that the coordinates $(q^i,p_j)$ are canonical is
equivalent with the following conditions
\begin{equation*}
\{Q^i,Q^j\} = \{P_i,P_j\} = 0, \quad \{Q^i,P_j\} = \delta^i_j.
\end{equation*}
From the time evolution equation \eqref{eq:7.1} we get Hamilton equations
\begin{align*}
\frac{\dd{Q^i}}{\dd{t}} = \{Q^i,H\} \quad \iff & \quad
\frac{\dd{q^i}}{\dd{t}} = \frac{\partial H}{\partial p_i}, \\
\frac{\dd{P_i}}{\dd{t}} = \{P_i,H\} \quad \iff & \quad
\frac{\dd{p_i}}{\dd{t}} = -\frac{\partial H}{\partial q_i}.
\end{align*}

In classical mechanics (to be compared with quantum mechanics) the uncertainty
relations for observables of position and momentum take the form
\begin{equation}
\Delta Q^i \Delta P_j \ge 0, \qquad i,j = 1,\dotsc,N,
\label{eq:7.3}
\end{equation}
where
\begin{equation*}
\Delta A = \sqrt{\braket{A^2}_\rho - \braket{A}^2_\rho}
\end{equation*}
is the uncertainty of an observable $A$ in a state $\rho$. Note that the
equality in \eqref{eq:7.3} takes place for pure states, thus in classical
mechanics pure states are simultaneously coherent states.

\section{Invariant quantization of Hamiltonian systems}
\label{sec:2}

\subsection{Preliminaries}
\label{subsec:2.1}
Let $(M,\mathcal{P},H)$ be a classical Hamiltonian system and $\mathcal{A}_C =
(C^\infty(M),\cdot,\{\sdot,\sdot\},\bar{\ })$ be a classical Poisson algebra.
By a quantization of such a Hamiltonian system we understand such a procedure
which modifies the classical uncertainty relations \eqref{eq:7.3} to quantum
(Heisenberg) uncertainty relations
\begin{equation*}
\Delta Q^i \Delta P_j \ge \frac{1}{2}\hbar\delta^i_j,
\end{equation*}
where $\hbar$ is a quantization parameter. The most natural quantization scheme
accomplishing this task is deformation quantization (see
\cite{Dito.Sternheimer:2002,Gutt:2000,Weinstein:1994b,Curtright:1998,%
Curtright:1999a,Blaszak:2012} for recent reviews). It allows a smooth passage
from classical to quantum theory, and also introduces quantization in a
geometric language similar to that of classical mechanics. Since the classical
mechanics can be formulated in a coordinate independent way the proper
quantization procedure should also be formulated in an invariant form. Below we
propose such a quantization scheme based on the deformation of classical
mechanics. Moreover, the invariant formulation of quantum mechanics will allow a
straightforward investigation of a quantum system in different coordinates.

The deformation quantization is based on a deformation of an algebraic structure
of the classical Poisson algebra $\mathcal{A}_C$ associated with the classical
Hamiltonian system. This will then yield a deformation of a
phase space (a Poisson manifold) to a noncommutative phase space
(a noncommutative Poisson manifold), a deformation of classical states to
quantum states and a deformation of classical observables to quantum
observables. The deformation is understood with respect to some parameter which
for physical reasons is taken to be the Planck's constant $\hbar$.
Moreover, in the limit $\hbar \to 0$ the quantum theory should reduce to the
classical theory.

In that process the classical Poisson algebra $\mathcal{A}_C =
(C^\infty(M),\cdot,\{\sdot,\sdot\},\bar{\ })$ is deformed to some
noncommutative algebra $\mathcal{A}_Q =
(C^\infty(M;\hbar),\star,\lshad\sdot,\sdot\rshad,*)$, where $\star$ is some
noncommutative associative product of functions being a deformation of
a point-wise product, $\lshad\sdot,\sdot\rshad$ is a Lie bracket satisfying
the Leibniz's rule and being a deformation of the Poisson bracket
$\{\sdot,\sdot\}$, and $*$ is an involution in the algebra $\mathcal{A}_Q$,
being a deformation of the classical involution (the complex-conjugation of
functions).

A deformation of a phase space is introduced as follows. A Poisson manifold
(phase space) $(M,\mathcal{P})$ is fully described by a Poisson algebra
$\mathcal{A}_C$. Hence by deforming $\mathcal{A}_C$ to a noncommutative algebra
$\mathcal{A}_Q$, we can think of a quantum Poisson algebra $\mathcal{A}_Q$ as
describing a noncommutative Poisson manifold.

The introduction of a global \nbr{\star}product for a general Poisson manifold
constitutes some problems, although Kontsevich proved that such a
\nbr{\star}product always exists \cite{Kontsevich:2003}. In this paper we are
most interested in a case when a phase space $M$ is induced by a
configuration space $\mathcal{Q}$ of the form of a pseudo-Euclidean space
$E^{r,s}$, i.e. $M = T^*E^{r,s}$. However, many results will apply
for more general phase spaces. In fact we can often assume that the Poisson
manifold $(M,\mathcal{P})$ is contractible to a point. As we will see later this
assumption guaranties, among other things, the existence of a global Darboux
coordinate system. Moreover, we will consider only Poisson tensors which are
non-degenerate.

Our quantization procedure is based on the observation that for a Poisson
manifold contractible to a point a Poisson tensor can be globally presented in
a form
\begin{equation}
\mathcal{P} = \sum_{i=1}^N X_i \wedge Y_i
= \sum_{i=1}^N (X_i \otimes Y_i - Y_i \otimes X_i),
\label{eq:2.1}
\end{equation}
for some pair-wise commuting vector fields $X_i,Y_i$ ($i = 1,\dotsc,N$), i.e.
$[X_i,Y_j] = [X_i,X_j] = [Y_i,Y_j] = 0$. The commutation of
the vector fields $X_i,Y_i$ guaranties the Jacobi identity for the Poisson
bracket induced by $\mathcal{P}$. We can now associate with such a Poisson
manifold the following natural \nbr{\star}product
\begin{align}
f \star g & = f \exp \left(
    \frac{1}{2} i\hbar\sum_k \overleftarrow{X}_k \wedge \overrightarrow{Y}_k
    \right) g \nonumber \\
& = f \exp \left(
    \frac{1}{2} i\hbar \sum_k\overleftarrow{X}_k \overrightarrow{Y}_k
    - \frac{1}{2} i\hbar \sum_k\overleftarrow{Y}_k \overrightarrow{X}_k
    \right) g \nonumber \\
& = f \cdot g + o(\hbar),
\label{eq:2.2}
\end{align}
where $X_k,Y_k$ are vector fields in the decomposition of the Poisson tensor
$\mathcal{P}$ according to \eqref{eq:2.1}. Then, a quantum Poisson bracket
$\lshad\sdot,\sdot\rshad$ is defined as follows
\begin{equation}
\lshad f,g \rshad_\star = \frac{1}{i\hbar} [f,g]
= \frac{1}{i\hbar}(f \star g - g \star f) = \{f,g\} + o(\hbar).
\label{eq:2.3}
\end{equation}

The \nbr{\star}product \eqref{eq:2.2} is the most natural one since it is
associative, has a particularly simple form, and the complex-conjugation is
the involution for this product:
\begin{equation*}
\overline{f \star g} = \bar{g} \star \bar{f}.
\label{eq:2.14}
\end{equation*}
Note moreover, that the associativity of the \nbr{\star}product, which follows
from the commutativity of the vector fields $X_i,Y_i$, guaranties that the
quantum Poisson bracket \eqref{eq:2.3} satisfies the Jacobi identity.
Furthermore, as we will see in \secref{sec:4} this \nbr{\star}product is related
to a symmetric (Weyl) ordering of position and momentum operators. It is
possible to introduce other \nbr{\star}products for which the
complex-conjugation will not be the involution, and which will be related to
different orderings (see \subsecref{subsec:2.7}).

Note that the \nbr{\star}product \eqref{eq:2.2} is not uniquely specified by a
Poisson manifold. The reason for this is that the representation \eqref{eq:2.1}
of the Poisson tensor as a wedge product of commuting vector fields is not
unique. There exist different commuting vector fields $X'_i,Y'_i$ giving the
same Poisson tensor. Thus to every Poisson manifold we can associate the whole
family of \nbr{\star}products \eqref{eq:2.2} parametrized by sequences of vector
fields $X_i,Y_i$ from the decomposition \eqref{eq:2.1} of the Poisson tensor. As
we will see later on, this family of \nbr{\star}products consists of equivalent
\nbr{\star}products.

\subsection{Equivalence of star-products}
\label{subsec:2.2}
Two star-products $\star$ and $\star'$ on a Poisson manifold $(M,\mathcal{P})$
are said to be equivalent if there exists a morphism
\begin{equation}
S = \id + \sum_{k=1}^\infty \hbar^k S_k,
\label{eq:2.5}
\end{equation}
where $S_k$ are linear operators on $C^\infty(M;\hbar)$, such that
\begin{equation}
S(f \star g) = Sf \star' Sg.
\label{eq:2.6}
\end{equation}

As was mentioned earlier, to every contractible Poisson manifold
$(M,\mathcal{P})$, whose Poisson tensor $\mathcal{P}$ can be written in the form
\eqref{eq:2.1}, corresponds a family of equivalent \nbr{\star}products of the
form \eqref{eq:2.2} parametrized by sequences of vector fields $X_i,Y_i$ from
the decomposition \eqref{eq:2.1} of the Poisson tensor. In other words, if
$\mathcal{P} =\sum_i X_i \wedge Y_i =\sum_i X'_i \wedge Y'_i$ and $\star$,
$\star'$ are star-products given by vector fields $X_i,Y_i$ and $X'_i,Y'_i$
respectively, then there exists a morphism $S$ of the form \eqref{eq:2.5}
satisfying \eqref{eq:2.6} \cite[Proposition~18]{Gutt:1999}.

\begin{example}
Let us consider the Poisson manifold $\mathbb{R}^2$ with the standard Poisson
tensor $\mathcal{P}$. Assume that $(x,p)$ is a Darboux coordinate system.
Consider the following vector fields
\begin{subequations}
\begin{gather*}
X = \partial_x, \quad Y = \partial_p, \\
X' = x^2\partial_x - 2xp\partial_p, \quad Y' = x^{-2}\partial_p.
\end{gather*}
\end{subequations}
It can be checked that $[X,Y] = 0$, $[X',Y'] = 0$ and
\begin{equation*}
\mathcal{P} = X \wedge Y = X' \wedge Y'.
\end{equation*}
Star-products induced by vector fields $X,Y$ and $X',Y'$ are equivalent and
the morphism $S$ giving this equivalence is represented by the formula
\begin{equation*}
S = \id + \frac{\hbar^2}{4}\left(2x^{-2}\partial_p^2 + x^{-2}p\partial_p^3
    - x^{-1}\partial_x\partial_p^2\right) + o(\hbar^4).
\end{equation*}
Note that vector fields $X,Y$ and $X',Y'$ are related by a canonical
transformation $T \colon (x,p) \mapsto T(x,p) = (-x^{-1},x^2 p)$:
\begin{equation*}
(Xf) \circ T = X'(f \circ T), \quad (Yf) \circ T = Y'(f \circ T),
\end{equation*}
for $f \in C^\infty(\mathbb{R}^2)$.
\end{example}

\subsection{Observables}
\label{subsec:2.4}
Similarly as in classical mechanics, in phase space quantum mechanics
observables are defined as functions from $C^\infty(M;\hbar)$, which are
self-adjoint with respect to the involution from $\mathcal{A}_Q$. To every
measurable quantity corresponds such function. However, different functions will
correspond to a given measurable quantity, depending on the chosen quantization.
In particular, quantum observables do not have to be the same functions as in
the classical case; they will be \nbr{\hbar}deformations of classical
observables. They do not even have to be real valued if the involution from
$\mathcal{A}_Q$ is not the complex-conjugation.

If $S$ is a morphism \eqref{eq:2.5} between two star-products $\star$ and
$\star'$, then it maps observables from one quantization scheme to the other.
Thus if $A$ is an observable corresponding to some measurable quantity in the
quantization scheme given by the \nbr{\star}product, then $A' = SA$ is an
observable in the quantization scheme given by the \nbr{\star'}product
corresponding to the same measurable quantity. In the limit $\hbar \to 0$ both
observables $A$ and $A'$ will reduce to the same classical observable.

Summarizing our previous considerations one observes that an explicit choice of
quantization of a classical Hamiltonian system is fixed by a choice of both,
the \nbr{\star}product and the form of quantum observables. In other words, one
needs to choose a particular deformation
of classical observables. It seems that there is no way of telling which
deformation of classical observables is appropriate for a given star-product
--- this can be only verified through experiment. On the other hand, there is a
very restrictive number of known physical quantum systems, being counterparts of
some classical systems. They are mainly described by so called natural
Hamiltonians with flat metrics (see \subsecref{subsec:4.2} and
\subsecref{subsec:4.3}). This knowledge is not enough to fix uniquely the
quantization and is the source of ambiguities. In consequence, the reader meets
in literature various versions of quantizations which coincide for the
class of natural flat Hamiltonians.

A phase space of a classical system over a configuration space $\mathcal{Q} =
E^{r,s}$ is equal: $M = T^*E^{r,s}$, with a Poisson tensor $\mathcal{P} =
\partial_{x^i} \wedge \partial_{p_i}$ for a pseudo-Euclidean coordinate system
$(x,p)$. In this paper we mainly focus on quantization of this kind of systems.
A canonical \nbr{\star}product corresponding to such phase space is a product
which in a pseudo-Euclidean coordinates $(x,p)$ has the form of the Moyal
product, i.e. the \nbr{\star}product for which $X_i = \partial_{x^i}$,
$Y_i = \partial_{p_i}$. It happens that for this \nbr{\star}product the choice
of quantum observables equal exactly to the classical observables is an
admissible natural choice, which leads to the standard Weyl quantization which
is consistent with the experiment. In the following sections we also make that
choice of quantization, but as was mentioned above there are other admissible
choices (see discussion in \secref{sec:5}).

\subsection{Examples of other star-products}
\label{subsec:2.7}
As was pointed out earlier, the \nbr{\star}product \eqref{eq:2.2} is not the
only \nbr{\star}product which can be defined. An example of a three-parameter
family of \nbr{\star}product equivalent with \eqref{eq:2.2}, in a case of a
two-dimensional phase space, is the following
\begin{align}
f \star_{\sigma,\alpha,\beta} g & = f \exp \Bigl(
    i\hbar (\tfrac{1}{2} - \sigma) \overleftarrow{X} \overrightarrow{Y}
    - i\hbar (\tfrac{1}{2} + \sigma) \overleftarrow{Y} \overrightarrow{X}
    \eqbreakadd + \hbar \alpha \overleftarrow{X} \overrightarrow{X}
    + \hbar \beta \overleftarrow{Y} \overrightarrow{Y} \Bigr) g,
\label{eq:2.8}
\end{align}
where $X,Y$ are commuting vector fields from the decomposition \eqref{eq:2.1} of
$\mathcal{P}$ and $\sigma,\alpha,\beta \in \mathbb{R}$. An isomorphism
\eqref{eq:2.5} intertwining the \nbr{\star_{\sigma,\alpha,\beta}}product with
the \nbr{\star}product \eqref{eq:2.2} reads
\begin{equation*}
S_{\sigma,\alpha,\beta} = \exp \left( -i\hbar \sigma XY
    + \tfrac{1}{2} \hbar \alpha XX + \tfrac{1}{2} \hbar \beta YY \right).
\label{eq:2.9}
\end{equation*}
The involution for the \nbr{\star_{\sigma,\alpha,\beta}}product takes the form
\begin{equation}
f^* = \exp \left( -2i\hbar \sigma XY \right) \bar{f}.
\label{eq:2.7}
\end{equation}
\Eqref{eq:2.7} indeed defines a proper involution. To see this first note that
the involution \eqref{eq:2.7} can be written in the form
$f^* = S_{\sigma,\alpha,\beta}\overline{S_{\sigma,\alpha,\beta}^{-1}f}$
\cite{Blaszak:2012}. Then from \eqref{eq:2.6} and the fact that the
complex-conjugation is the involution for the \nbr{\star}product we get
\begin{align*}
(f \star_{\sigma,\alpha,\beta} g)^* & = S_{\sigma,\alpha,\beta}
    \overline{S_{\sigma,\alpha,\beta}^{-1}(f \star_{\sigma,\alpha,\beta} g)}
    \nonumber \\
& = S_{\sigma,\alpha,\beta}(\overline{S_{\sigma,\alpha,\beta}^{-1}f \star
    S_{\sigma,\alpha,\beta}^{-1}g}) \nonumber \\
& = S_{\sigma,\alpha,\beta}(\overline{S_{\sigma,\alpha,\beta}^{-1}g} \star
    \overline{S_{\sigma,\alpha,\beta}^{-1}f}) \nonumber \\
& = (S_{\sigma,\alpha,\beta}\overline{S_{\sigma,\alpha,\beta}^{-1}g})
    \star_{\sigma,\alpha,\beta}
    (S_{\sigma,\alpha,\beta}\overline{S_{\sigma,\alpha,\beta}^{-1}f})
    \nonumber \\
& = g^* \star_{\sigma,\alpha,\beta} f^*.
\end{align*}

From \eqref{eq:2.7} it is evident that for $\sigma \neq 0$ the involution for
the \nbr{\star_{\sigma,\alpha,\beta}}product is different from the
complex-conjugation and functions self-adjoint with respect to it can be in
general complex.

\begin{example}
\label{eg:2.2}
As an example let us consider the quantization $\star_{\sigma,\alpha,\beta}$
\eqref{eq:2.8} in a natural coordinate system when $X = \partial_x$ and
$Y = \partial_p$. Consider complex function $A(x,p) = xp^2 + \hbar\beta x
- 2i\hbar\sigma p$. A simple calculation shows that $A$ represents an
observable, as it is self-adjoint with respect to the involution $*$
\eqref{eq:2.7}. Moreover, it is equivalent to observable $A=xp^2$ for Moyal
quantization in the same coordinate system.
\end{example}

With the \nbr{\star}product \eqref{eq:2.8} there are associated orderings of the
position and momentum operators different than the symmetric ordering
\cite{Blaszak:2012}. For example, for the cases $\sigma = \pm \frac{1}{2}$,
$\alpha = \beta = 0$ correspond normal and anti-normal orderings.

In literature it is common to find a situation when one uses different
orderings, for example normal ordering, when quantizing a classical system,
without any change of classical observables when constructing from them
operators. This leads, in general, to operators which are not Hermitian.

Further discussion about admissible \nbr{\star}products is presented in
\subsecref{subsec:5.2}.

\subsection{Coordinate systems}
\label{subsec:2.3}
Poisson manifolds as well as \nbr{\star}product defined on them can be
investigated in different coordinate systems. Let $X_i,Y_i$ be vector fields
from the definition \eqref{eq:2.2} of a \nbr{\star}product. On every
contractible manifold there exists a global coordinate system $(x,p)$ in which
$X_i,Y_i$ are coordinate vector fields, i.e. $X_i = \partial_{x^i}$,
$Y_i = \partial_{p_i}$. Such a coordinate system is of course a Darboux
coordinate system associated with the Poisson tensor $\mathcal{P}$. It will be
called a natural coordinate system for the star-product. In these coordinates
the \nbr{\star}product takes the form of a Moyal product.

Note, that if $\star$ and $\star'$ are two star-products on the Poisson manifold
$(M,\mathcal{P})$, of the form \eqref{eq:2.2}, and $(x,p)$ and $(x',p')$ natural
coordinates associated with them, then a transformation $(x,p) \mapsto (x',p')$
is a classical canonical transformation.

If $(x,p)$ is some arbitrary quantum canonical coordinate system on the phase
space, then functions $Q^i(x,p) = x^i$ and $P_j(x,p) = p_j$ are observables of
position and momentum for this coordinate system. The quantum canonicity of a
coordinate system means that
\begin{equation*}
\lshad Q^i,Q^j \rshad = \lshad P_i,P_j \rshad = 0, \quad
\lshad Q^i,P_j \rshad = \delta^i_j.
\end{equation*}
If, moreover, there holds
\begin{equation*}
\{Q^i,Q^j\} = \{P_i,P_j\} = 0, \quad \{Q^i,P_j\} = \delta^i_j,
\end{equation*}
then $(x,p)$ is also a classical canonical coordinate system. Note that a
natural coordinate system for a \nbr{\star}product is a classical canonical, as
well as a quantum canonical coordinate system.

In what follows we will use the following notation. A \nbr{\star}product
\eqref{eq:2.2} written in a coordinate system $(x,p)$ will be denoted by
$\star^{(x,p)}$. In addition $\star_M^{(x,p)}$ will denote a \nbr{\star}product
written in $(x,p)$ coordinates, which has the form of the Moyal product.

A crucial observation important when dealing with operator quantum mechanics is
that the quantization given by a star-product $\star$ and considered in some
quantum canonical coordinate system $(x,p)$ is equivalent with a quantization
for which a star-product in the coordinates $(x,p)$ has the form of the Moyal
product $\star_M^{(x,p)}$. Moreover, this equivalence can be chosen so that it
will preserve the observables of position and momentum (for both quantizations
to observables of position and momentum will correspond the same functions).
Furthermore, the equivalence with such a property is specified uniquely. In
other words there exists a unique isomorphism $S$ of the form \eqref{eq:2.5}
satisfying
\begin{equation}
S(f \star_M^{(x,p)} g) = Sf \star^{(x,p)} Sg
\label{eq:2.12}
\end{equation}
and
\begin{gather}
SQ^i = Q^i, \quad SP_j = P_j, \label{eq:2.10} \\
\overline{S(f)} = S(\bar{f}). \label{eq:2.11}
\end{gather}
The proof of the existence and uniqueness of such an isomorphism, together with
a systematic construction of it, in a case of a coordinate system classical and
quantum canonical, is given in \cite{Domanski:2013}. We believe that such an
isomorphism also exists for a general quantum canonical transformation, not
necessarily classical canonical.

If the coordinate system $(x,p)$ is not quantum canonical but only classical
canonical, then the \nbr{\star}product written in it will also be equivalent
with a Moyal product. However, the relation \eqref{eq:2.10} will not have to be
satisfied anymore. In such a case, as it is with other observables, the
observables of position and momentum $Q^i$, $P_j$ have to be deformed to
functions $\tilde{Q}^i = S^{-1}Q^i$, $\tilde{P}_j = S^{-1}P_j$. The functions
$\tilde{Q}^i$, $\tilde{P}_j$ are proper observables of position and momentum
satisfying
\begin{equation*}
\lshad \tilde{Q}^i,\tilde{Q}^j \rshad = \lshad \tilde{P}_i,\tilde{P}_j \rshad
= 0, \quad \lshad \tilde{Q}^i,\tilde{P}_j \rshad = \delta^i_j.
\end{equation*}
They also define a quantum canonical coordinate system.

Let us consider a transformation of coordinates on a configuration space from
a pseudo-Euclidean coordinates $(x^1,\dotsc,x^N)$ to a new coordinate system
$(x'^1,\dotsc,x'^N)$: $\phi \colon V \subset \mathbb{R}^N \to U \subset
\mathbb{R}^N$, $x = \phi(x')$. Such a transformation induces a classical
canonical transformation on a phase space $T(x',p') = (x,p)$, where
\begin{subequations}
\label{eq:2.13}
\begin{align}
x^i & = \phi^i(x'), \label{eq:2.13a} \\
p_i & = [(\phi'(x'))^{-1}]^j_i p'_j, \label{eq:2.13b}
\end{align}
\end{subequations}
and $[(\phi'(x'))^{-1}]^j_i$ denotes an inverse matrix to the Jacobian matrix
$[\phi'(x')]^i_j = \frac{\partial \phi^i}{\partial x^j}(x')$ of $\phi$. The
transformation $T$ is called a point transformation. A simple
calculation shows that it is also quantum canonical. A Moyal product
$\star_M^{(x,p)}$ transformed by the transformation $T$ takes the form
\begin{equation*}
f \star^{(x',p')} g = f \exp \left(
    \frac{1}{2}i\hbar \overleftarrow{D}_{x'^i} \overrightarrow{D}_{p'_i}
    - \frac{1}{2}i\hbar \overleftarrow{D}_{p'_i} \overrightarrow{D}_{x'^i}
    \right) g,
\end{equation*}
where
\begin{subequations}
\begin{align*}
D_{x'^i} & = [(\phi'(x'))^{-1}]^j_i \partial_{x'^j} \eqbreakadd
    + [(\phi'(x'))^{-1}]^j_i [(\phi'(x'))^{-1}]^r_k [\phi''(x')]^k_{jl}
    p'_r \partial_{p'_l}, \\
D_{p'_i} & = [\phi'(x')]^i_j \partial_{p'_j},
\end{align*}
\end{subequations}
and $[\phi''(x')]^i_{jk} = \frac{\partial^2 \phi^i}{\partial_{x'^j}
\partial_{x'^k}}(x')$ is the Hessian of $\phi$. The isomorphism \eqref{eq:2.12}
intertwining the \nbr{\star^{(x',p')}}product with the Moyal product
$\star_M^{(x',p')}$ written in the coordinates $(x',p')$ takes the following
form, up to the second order in $\hbar$
\begin{align}
S_T & = \id + \frac{\hbar^2}{4!} \Bigl(3\Gamma^i_{lj}(x')\Gamma^l_{ik}(x')
    \partial_{p'_j}\partial_{p'_k}
    + 3\Gamma^i_{jk}(x') \partial_{x'^i}\partial_{p'_j}\partial_{p'_k}
    \nonumber \\
& \quad {} + \left( 2\Gamma^i_{nl}(x')\Gamma^n_{jk}(x')
- \partial_{x'^l}\Gamma^i_{jk}(x') \right)
    p'_i \partial_{p'_j}\partial_{p'_k}\partial_{p'_l} \Bigr) \eqbreakadd
    + o(\hbar^4),
\label{eq:2.17}
\end{align}
where
\begin{equation}
\Gamma^i_{jk}(x') = [(\phi'(x'))^{-1}]^i_r [\phi''(x')]^r_{jk}.
\label{eq:2.16}
\end{equation}
Note that the symbols $\Gamma^i_{jk}(x')$ are the Christoffel symbols for the
$(x'^1,\dotsc,x'^N)$ coordinates, associated to the Levi-Civita connection
$\nabla$ on the configuration space $\mathcal{Q} = E^{r,s}$.

Let us make some remarks about domains of coordinate systems. If one is
interested only in the investigation of a geometry of a classical Hamiltonian
system $(M,\mathcal{P},H)$, then one can consider coordinate systems defined on
arbitrary open subsets $U$ of a phase space $M$. The same thing is true for
quantum systems considered in the framework of the deformation quantization,
since one can easily restrict a star-product to a space of functions
$C^\infty(U)$ defined on an open subset $U$ of the phase space $M$.

However, when one wishes to investigate integrals over the phase space, e.g., to
calculate expectation values of observables, then one cannot do this in an
arbitrary coordinate system. The reason for this is that, in general the values
of integrals will change if the integration will be performed over some subset
$U \subset M$. This argument applies both to classical and quantum theory.
The only coordinate systems in which it is meaningful to consider integration
are those which are defined on almost the whole phase space, i.e. on an open
subset $U \subset M$ such that $M \setminus U$ is a set of Liouville-measure
zero. Such coordinate systems do not change integrals.

As an example let us try to calculate the expectation value of an observable $A$
in a state $\rho$ in a polar coordinate system. Assume that $\braket{A}_\rho$ is
given in a pseudo-Euclidean coordinate system. The transformation
to the polar coordinate system is a map $T \colon V \times \mathbb{R}^2 \to
U \times \mathbb{R}^2$, where $V = (0,\infty) \times [0,2\pi)$,
$U = \mathbb{R}^2 \setminus \{0\}$, $T(r,\theta,p_r,p_\theta) = (x,y,p_x,p_y)$.
Note that $\mathbb{R}^2 \setminus U$ is of measure zero. Thus we have
\begin{align}
\braket{A}_\rho & = \int_{\mathbb{R}^2} \int_{\mathbb{R}^2} A(x,y,p_x,p_y)
    \rho(x,y,p_x,p_y) \ud{x}\ud{y}\ud{p_x}\ud{p_y} \nonumber \\
& = \int_U \int_{\mathbb{R}^2} A(x,y,p_x,p_y) \rho(x,y,p_x,p_y)
    \ud{x}\ud{y}\ud{p_x}\ud{p_y} \nonumber \\
& = \int_V \int_{\mathbb{R}^2} A(T(r,\theta,p_r,p_\theta))
    \rho(T(r,\theta,p_r,p_\theta)) \eqbreakmul
    \ud{r}\ud{\theta}\ud{p_r}\ud{p_\theta}.
\end{align}

Another problem appears when one wishes to pass to the Hilbert space approach
of quantum mechanics. This passage cannot be done in an arbitrary coordinate
system. To see this assume that a quantum system is described, in the
pseudo-Euclidean coordinates, by the Hilbert space $L^2(\mathbb{R}^N)$. In
some other coordinates the Hilbert space describing the quantum system could be
$L^2(V,\mu)$, where $V \subset \mathbb{R}^N$ is some open subset and $\mu$ is
some integration measure. One could now pass to a representation corresponding
to the pseudo-Euclidean coordinate system receiving a unitary operator mapping
the Hilbert space $L^2(V,\mu)$ onto a Hilbert space $L^2(U)$, where
$U \subset \mathbb{R}^N$ is some open subset. However, the Hilbert spaces
$L^2(\mathbb{R}^N)$ and $L^2(U)$, in general, will describe two non-equivalent
quantum systems, despite the fact that they should describe the same quantum
system in the pseudo-Euclidean coordinates. Again the only possible coordinate
systems, in which one can pass to Hilbert space approach of quantum mechanics,
are those defined on almost the whole phase space. For such coordinate systems,
when $\mathbb{R}^N \setminus U$ is of measure zero, the Hilbert spaces
$L^2(\mathbb{R}^N)$ and $L^2(U)$ are naturally isomorphic.

\subsection{Quantum states}
\label{subsec:2.5}
Let us assume that the phase space $M = T^*E^{r,s}$ and that the
quantization is given by the star-product of the form \eqref{eq:2.2}. For such
quantum system states can be defined, in analogy with classical theory, as
square integrable functions $\rho$ defined on the phase space $M$ satisfying
the following conditions:
\begin{enumerate}
\item  $\rho = \bar{\rho}$ (self-conjugation),
\item  $\displaystyle \int_M \rho \ud{\Omega} = 1$ (normalization),
\item  $\displaystyle \int_M \bar{f} \star f \star \rho \ud{\Omega} \ge 0$ for
$f \in C^\infty(M;\hbar)$ (positive define).
\end{enumerate}
Quantum states form a convex subset of the Hilbert space $L^2(M)$. For this
reason the Hilbert space $\mathcal{H} = L^2(M)$ of square integrable functions
on the phase space will be called a space of states. Observe, that in the
definition of states the fact that the \nbr{\star}product can be extended to a
product between smooth functions from $C^\infty(M;\hbar)$ and square integrable
functions from $L^2(M)$ was used. It is also possible to define the
\nbr{\star}product between square integrable functions from $L^2(M)$ by
extending it from the space $\mathcal{S}(M)$ of Schwartz functions
\cite{Blaszak:2012}. Note also that quantum states are closed with respect to
the \nbr{\star}product.

Pure states are defined as those states which cannot be written as convex linear
combinations of some other states, i.e., $\rho_{\text{pure}}$ is a pure state
if and only if there do not exist two different states $\rho_1$ and $\rho_2$
such that $\rho_{\text{pure}} = p\rho_1 + (1 - p)\rho_2$ for some
$p \in (0,1)$. A state which is not pure is called a mixed state.

Pure states can be alternatively characterized as functions
$\rho_{\text{pure}} \in \mathcal{H}$ which are idempotent (compare with
classical case \eqref{eq:7.5}):
\begin{equation*}
\rho_{\text{pure}} \star \rho_{\text{pure}} =
\frac{1}{(2\pi\hbar)^N} \rho_{\text{pure}}.
\end{equation*}
Mixed states $\rho_{\text{mix}} \in \mathcal{H}$ can be characterized as convex
linear combinations, possibly infinite, of some families of pure states
$\rho_{\text{pure}}^{(\lambda)}$
\begin{equation*}
\rho_{\text{mix}} = \sum_{\lambda} p_{\lambda} \rho_{\text{pure}}^{(\lambda)},
\end{equation*}
where $p_{\lambda} \ge 0$ and $\sum_{\lambda} p_{\lambda} = 1$.

For a given observable $A \in C^\infty(M;\hbar)$ and state $\rho$ the
expectation value of the observable $A$ in the state $\rho$ is defined by
\begin{equation}
\braket{A}_\rho := \int_M A \star \rho \ud{\Omega}
= \int_M A \cdot \rho \ud{\Omega}.\label{eq:2.30}
\end{equation}
The last equality in \eqref{eq:2.30} is valid only for \nbr{\star}products of
the form \eqref{eq:2.2}.

\subsection{Time evolution of quantum Hamiltonian systems}
\label{subsec:2.6}
The time evolution of a quantum system is governed by a Hamilton function $H$
which is, similarly as in classical mechanics, some distinguished observable.
As in classical theory there are two dual points of view on the time evolution:
Schr\"odinger picture and Heisenberg picture. In the Schr\"odinger picture
states undergo time development while observables do not. An equation of motion
for states, through an analogy to Liouville equation, takes the form
\begin{equation}
\frac{\partial \rho}{\partial t}(t) - \lshad H,\rho(t) \rshad = 0.
\label{eq:2.15}
\end{equation}
In the Heisenberg picture states remain still whereas observables undergo the
time development. A time evolution equation for observables, through an analogy
to the classical case, reads
\begin{equation*}
\frac{\dd{A}}{\dd{t}}(t) - \lshad A(t),H \rshad = 0.
\end{equation*}
Both presented approaches to the time development yield equal predictions
concerning the results of measurements, since
\begin{equation*}
\braket{A(0)}_{\rho(t)} = \braket{A(t)}_{\rho(0)}.
\end{equation*}

\section{Operator representation over flat configuration space of quantum
mechanics}
\label{sec:4}
Let us consider a classical system described by a phase space $M = T^*E^{r,s}$,
$\mathcal{P} = \partial_{x^i} \wedge \partial_{p_i}$, and its canonical
quantization, i.e. a quantization given by a star-product \eqref{eq:2.2} such
that in the pseudo-Euclidean coordinates it takes the form of the Moyal
product and quantum observables are equal the classical ones. The passage to a
standard approach to quantum mechanics, i.e. an operator representation where a
Hilbert space of states is represented by a space of square integrable functions
over the configuration space, has to be performed in some coordinate system.
First let us choose the pseudo-Euclidean coordinate system.

\subsection{Quantum mechanics in a pseudo-Euclidean coordinate system}
\label{subsec:4.1}
Let $(x,p)$ be a pseudo-Euclidean coordinate system. In these coordinates the
star-product takes the form of the Moyal product. Now, note that the Hilbert
space of states $\mathcal{H} = L^2(M)$ in these coordinates is equal
$L^2(\mathbb{R}^{2N})$ and can be written as the following tensor product of
the Hilbert space $L^2(\mathbb{R}^N)$ and a space dual to it
$(L^2(\mathbb{R}^N))^*$:
\begin{equation*}
\mathcal{H} = \left(L^2(\mathbb{R}^N)\right)^* \otimes_M L^2(\mathbb{R}^N),
\end{equation*}
where the tensor product $\otimes_M$ is defined by
\begin{align*}
(\varphi^* \otimes_M \psi)(x,p) & = \frac{1}{(2\pi \hbar)^{N/2}} \int \vd{y}
    e^{-\frac{i}{\hbar} py} \bar{\varphi}\left(x - \frac{1}{2} y\right)
    \eqbreakmul \psi\left(x + \frac{1}{2} y\right),
\end{align*}
where $\varphi, \psi \in L^2(\mathbb{R}^N)$. The Hilbert space
$L^2(\mathbb{R}^N)$ is the space of states for the standard approach to quantum
mechanics in a position representation corresponding to the coordinate system
$(x,p)$. Any $\rho_{\text{pure}}\in L^2(\mathbb{R}^{2N})$ is represented by
$\varphi \in L^2(\mathbb{R}^N)$ through the relation
$\rho_{\text{pure}}(x,p) = (\varphi^* \otimes_M \varphi)(x,p)$ being a well
known Wigner function.

States $\rho \in \mathcal{H}$ treated as operators $\hat{\rho} =
(2\pi\hbar)^N \rho \star_M^{(x,p)} {}$ can be written in the following form
\cite{Blaszak:2012}
\begin{equation}
\hat{\rho} = \hat{1} \otimes_M \hat{\varrho},
\label{eq:4.1}
\end{equation}
where $\hat{\varrho}$ is some density operator representing a state in the
standard approach to quantum mechanics. Hence to every pure or mixed
state $\rho \in \mathcal{H}$ corresponds a unique density operator
$\hat{\varrho}$. Similarly, observables $A \in \mathcal{A}_Q$ treated as
operators $\hat{A} = A \star_M^{(x,p)} {}$ take the form \cite{Blaszak:2012}
\begin{equation}
\hat{A} = A \star_M^{(x,p)} {} = \hat{1} \otimes_M A_W(\hat{q},\hat{p}),
\label{eq:4.2}
\end{equation}
where
\begin{equation}
A_W(\hat{q},\hat{p}) = A(-i\hbar \partial_\xi, i\hbar \partial_\eta)
    e^{\frac{i}{\hbar}(\xi_i \hat{q}^i - \eta^i \hat{p}_i)} \bigg|_{\xi=\eta=0}
\end{equation}
is the function $A$ of symmetrically ordered (Weyl ordered) operators of
position and momentum $\hat{q}^i = x^i$ and $\hat{p}_j = -i\hbar\partial_{x^j}$.
In particular, from this it follows that
\begin{subequations}
\label{eq:4.3}
\begin{align}
A \star_M^{(x,p)} \Psi & = \varphi^* \otimes_M A_W(\hat{q},\hat{p}) \psi, \\
\Psi \star_M^{(x,p)} A & = (A_W^\dagger(\hat{q},\hat{p})\varphi)^* \otimes_M
    \psi,
\end{align}
\end{subequations}
for $\Psi = \varphi^* \otimes_M \psi$ and $\varphi,\psi \in L^2(\mathbb{R}^N)$.

The expectation values of observables $A \in \mathcal{A}_Q$ in states $\rho
\in \mathcal{H}$ are the same as when computed in ordinary quantum mechanics
\begin{equation}
\braket{A}_{\rho} = \tr(\hat{\varrho} A_W(\hat{q},\hat{p})),
\label{eq:4.4}
\end{equation}
where $\hat{\varrho}$ is a density operator corresponding to $\rho$ and
$A_W(\hat{q},\hat{p})$ is an operator corresponding to $A$. Also the time
evolution equation \eqref{eq:2.15} of states $\rho \in \mathcal{H}$ corresponds
to the von Neumann equation describing the time evolution of density operators
$\hat{\varrho}$:
\begin{equation}
i\hbar\frac{\partial \hat{\varrho}}{\partial t}
    - [H_W(\hat{q},\hat{p}),\hat{\varrho}] = 0.
\label{eq:4.6}
\end{equation}

\subsection{Quantum mechanics in arbitrary coordinates on the configuration
space}
\label{subsec:4.2}
In the previous subsection we received a position representation of quantum
mechanics in the Hilbert space $L^2(\mathbb{R}^N)$ for observables of position
$\hat{q}^i$ corresponding to a pseudo-Euclidean coordinate system, i.e. we
reconstructed the Weyl quantization procedure. Let us choose some arbitrary
coordinate system $(x'^1,\dotsc,x'^N)$ on the configuration space. We will show
how to construct position operators $\hat{q}'^i$ corresponding to this
coordinate system and represent a quantum system in a position representation
corresponding to this new set of position observables.

Let
\begin{equation}
\phi \colon \mathbb{R}^{N} \supset V \to U \subset \mathbb{R}^{N}, \quad
x = \phi(x')
\label{eq:4.16}
\end{equation}
be a transformation from $(x'^1,\dotsc,x'^N)$ coordinates to a
pseudo-Euclidean coordinates, such that $\mathbb{R}^N \setminus U$ is of
measure zero. The transformation $\phi$ induces a classical and quantum
canonical transformation $T$ according to \eqref{eq:2.13}. Note that the
Hilbert spaces $L^2(\mathbb{R}^N)$ and $L^2(U)$ are naturally isomorphic and can
be identified with each other. We can define operators of position and momentum
corresponding to the coordinate system $(x',p')$ according to
\begin{subequations}
\begin{align*}
\hat{Q}^i & = (Q^i)_W(\hat{q}), \\
\hat{P}_j & = (P_j)_W(\hat{q},\hat{p}),
\end{align*}
\end{subequations}
where $T^{-1}(x,p) = (Q^1(x),\dotsc,P_N(x,p))$. These operators are defined on
the Hilbert space $L^2(U) \cong L^2(\mathbb{R}^N)$ corresponding to the
pseudo-Euclidean coordinate system. We can now use the operators $\hat{Q}^i$ to
create a position representation of the quantum system, corresponding to the
coordinates $(x',p')$. It follows that for a unitary operator
$\hat{U}_T \colon L^2(\mathbb{R}^N) \to L^2(V,\mu)$, where
$\dd{\mu(x')} = \abs{\phi(x')} \ud{x'} = \abs{\det[g'_{ij}(x')]}^{1/2} \ud{x'}$,
given by
\begin{equation}
(\hat{U}_T \varphi)(x') = \varphi(\phi(x')),
\label{eq:4.15}
\end{equation}
the following formula holds
\begin{equation*}
\hat{U}_T \hat{Q}^i \hat{U}_T^{-1} = x'^i \equiv \hat{q}'^i.
\end{equation*}
We thus have that the quantum system written in the position representation
corresponding to the coordinates $(x',p')$ is described by the Hilbert space
$L^2(V,\mu)$ and that the unitary operator $\hat{U}_T$ intertwines between two
representations corresponding to coordinates $(x,p)$ and $(x',p')$. Note that
the momentum operators $\hat{p}'_j = \hat{U}_T \hat{P}_j \hat{U}_T^{-1}$
corresponding to the coordinates $(x',p')$ and defined on the Hilbert space
$L^2(V,\mu)$ are given by the formulas
\begin{equation}
\hat{p}'_j = -i\hbar\left(\partial_{x'^j} + \frac{1}{2}\Gamma^k_{jk}(x')\right),
\label{eq:4.8}
\end{equation}
where $\Gamma^i_{jk}(x')$ are given by \eqref{eq:2.16} (see also
\cite{DeWitt:1952,Essen:1978}).
Indeed, $P_j(x,p) = p'_j = p_i [\phi'(\phi^{-1}(x))]^i_j$. Hence
\begin{align*}
(\hat{p}'_j\psi)(x') & = \left(\hat{U}_T (P_j)_W(\hat{q},\hat{p})\hat{U}_T^{-1}
    \psi\right)(x')
= \left( \hat{U}_T \left( \frac{1}{2} \hat{p}_i
    \frac{\partial \phi^i}{\partial x'^j}(\phi^{-1}(\hat{q}))
+ \frac{1}{2}\frac{\partial \phi^i}{\partial x'^j}(\phi^{-1}(\hat{q}))\hat{p}_i
    \right) \hat{U}_T^{-1}\psi\right)(x') \nonumber \\
& = -\frac{1}{2}i\hbar\partial_{x^i} \left(\frac{\partial \phi^i}{\partial x'^j}
    \circ \phi^{-1}\right)(\phi(x')) \psi(x')
- \frac{\partial \phi^i}{\partial x'^j}(x')
    \frac{1}{2}i\hbar\partial_{x^i}(\psi \circ \phi^{-1})(\phi(x')) \nonumber \\
& = -i\hbar\left( \frac{\partial \phi^i}{\partial x'^j}(x')
    \frac{\partial (\phi^{-1})^k}{\partial x^i}(\phi(x'))
    \frac{\partial \psi}{\partial x'^k}(x')
+ \frac{1}{2} \frac{\partial^2 \phi^i}{\partial x'^j \partial x'^k}(x')
    \frac{\partial (\phi^{-1})^k}{\partial x^i}(\phi(x')) \psi(x') \right).
\end{align*}
Using the identity $(\phi'(x'))^{-1} = (\phi^{-1})'(\phi(x'))$, from which
follows that $\frac{\partial (\phi^{-1})^k}{\partial x^i}(\phi(x')) =
[(\phi'(x'))^{-1}]^k_i$, we receive the result.

In \subsecref{subsec:4.1} we constructed the operator representation of the
quantum system written in pseudo-Euclidean coordinates. In what follows we
will show how to construct such representation for the quantum system written in
arbitrary coordinates. In fact the whole construction is similar to that for the
pseudo-Euclidean coordinates. The only difference is in that the tensor product
$\otimes_M$ has to be replaced with some other product and the Weyl ordering of
operators $\hat{q}^i$, $\hat{p}_j$ with some other ordering. Moreover, instead
of the Hilbert space $L^2(\mathbb{R}^N)$ the Hilbert space $L^2(V,\mu)$ has to
be used.

To find the form of the twisted tensor product and the ordering of operators
$\hat{q}^i$, $\hat{p}_j$ for an arbitrary coordinate system we
can use the fact that the star-product in these coordinates is equivalent to
the Moyal product (see \subsecref{subsec:2.3}). Let $S$ denote an isomorphism
giving this equivalence. Then the twisted tensor product, denoted by
$\otimes_S$, can be defined by the formula
\begin{equation*}
\varphi^* \otimes_S \psi := S(\varphi^* \otimes_M \psi),
\end{equation*}
and the new \nbr{S}ordering by the formula
\begin{equation*}
A_S(\hat{q},\hat{p}) := (S^{-1}A)_W(\hat{q},\hat{p}).
\end{equation*}
Formulas \eqref{eq:4.1}--\eqref{eq:4.6} hold true for a general quantum
canonical coordinate system, provided that we replace the tensor product
$\otimes_M$ with $\otimes_S$ and the symmetric ordering with \nbr{S}ordering
\cite{Blaszak:2012}.

The unitary operator $\hat{U}_T$ gives the equivalence of quantizations
performed in different coordinate systems as can be seen from the following
equality
\begin{equation}
(\varphi^* \otimes_M \psi) \circ T =
    (\hat{U}_T \varphi)^* \otimes_S \hat{U}_T \psi, \quad
    \varphi,\psi \in L^2(\mathbb{R}^N).
\label{eq:4.5}
\end{equation}
From \eqref{eq:4.5} it follows that operators, corresponding to a function
$A \in C^\infty(\mathbb{R}^{2N})$ written in different coordinate systems, are
unitary equivalent:
\begin{equation*}
A'_S(\hat{q}', \hat{p}') = (S^{-1}A')_W(\hat{q}', \hat{p}')
= \hat{U}_T A_W(\hat{q}, \hat{p}) \hat{U}_T^{-1},
\label{eq:4.7}
\end{equation*}
where $A' = A \circ T$.

\begin{example}
Let us consider a point transformation generated by a transformation to
spherical polar coordinate system $T(r,\theta,\phi,p_r,p_\theta,p_\phi) =
(x,y,z,p_x,p_y,p_z)$
\begin{subequations}
\begin{align*}
x & = r\sin\theta\cos\phi, \\
y & = r\sin\theta\sin\phi, \\
z & = r\cos\theta, \\
p_x & = \frac{rp_r \sin^2\theta\cos\phi + p_\theta \sin\theta\cos\theta\cos\phi
    - p_\phi \sin\phi}{r\sin\theta}, \\
p_y & = \frac{rp_r \sin^2\theta\sin\phi + p_\theta \sin\theta\cos\theta\sin\phi
    + p_\phi \cos\phi}{r\sin\theta}, \\
p_z & = \frac{rp_r \cos\theta - p_\theta \sin\theta}{r}.
\end{align*}
\end{subequations}
The isomorphism $S_T$ \eqref{eq:2.17} associated to this transformation takes
the form
\begin{align*}
S_T & = \id + \frac{\hbar^2}{4} \biggl( \frac{1}{r^2} \partial_{p_r}^2
+ \left(\frac{1}{2\tan^2\theta} - 1\right) \partial_{p_\theta}^2
- \partial_{p_\phi}^2
+ \frac{1}{r\tan\theta} \partial_{p_r} \partial_{p_\theta}
+ \frac{1}{r^2} p_\theta \partial_{p_r}^2 \partial_{p_\theta}
- \frac{1}{2} p_r \partial_{p_r} \partial_{p_\theta}^2 \nonumber \\
& \quad {} + \frac{2}{r\tan\theta} p_\phi \partial_{p_r} \partial_{p_\theta}
    \partial_{p_\phi}
- \left(\frac{1}{2} p_r \sin^2\theta + \frac{1}{r} p_\theta \sin\theta\cos\theta
    \right) \partial_{p_r} \partial_{p_\phi}^2
- \frac{1}{3} p_\theta \partial_{p_\theta}^3
+ \frac{1}{\tan^2\theta} p_\phi \partial_{p_\theta}^2 \partial_{p_\phi}
\nonumber \\
& \quad {} - \frac{1}{2} p_\theta \partial_{p_\theta} \partial_{p_\phi}^2
- \frac{1}{3} p_\phi \partial_{p_\phi}^3
+ \frac{1}{r^2} p_\phi \partial_r^2 \partial_{p_\phi}
- \frac{1}{2} r \partial_r \partial_{p_\theta}^2
- \frac{1}{2} r\sin^2\theta \partial_r \partial_{p_\phi}^2
+ \frac{1}{r} \partial_\theta \partial_{p_r} \partial_{p_\theta} \nonumber \\
& \quad {} - \frac{1}{2}\sin\theta\cos\theta \partial_\theta \partial_{p_\phi}^2
+ \frac{1}{r} \partial_\phi \partial_{p_r} \partial_{p_\phi}
+ \frac{1}{\tan\theta} \partial_\phi \partial_{p_\theta} \partial_{p_\phi}
\biggr) + o(\hbar^4).\label{eq:4.100}
\end{align*}

A quantum system after transformation to spherical coordinates will be described
by a Hilbert space $L^2(V,\mu)$, where
$V = (0,\infty) \times (0,\pi) \times (0,2\pi)$ and
$\dd{\mu(r,\theta,\phi)} = r^2\sin\theta \ud{r}\ud{\theta}\ud{\phi}$.
$L^2(V,\mu)$ is the Hilbert space of square integrable functions
defined on $V$.

The momentum operators associated to the spherical coordinate system take
the form
\begin{subequations}
\begin{align*}
\hat{p}_r & = -i\hbar\left(\partial_r + \frac{1}{r}\right), \\
\hat{p}_\theta & = -i\hbar\left(\partial_\theta+\frac{1}{2\tan\theta}\right), \\
\hat{p}_\phi & = -i\hbar\partial_\phi.
\end{align*}
\end{subequations}

Let us now consider a Hamiltonian $H$ of a hydrogen atom. In the Cartesian
coordinate system it takes the form
\begin{align*}
H(x,y,z,p_x,p_y,p_z) & = \frac{p_x^2 + p_y^2 + p_z^2}{2m} \eqbreakadd
    - \frac{1}{4\pi\epsilon_0} \frac{e^2}{\sqrt{x^2 + y^2 + z^2}}.
\end{align*}
In the spherical coordinates it can be written in the form
\begin{align*}
H'(r,\theta,\phi,p_r,p_\theta,p_\phi) & = \frac{1}{2m}\left(p_r^2
    + \frac{p_\theta^2}{r^2} + \frac{p_\phi^2}{r^2 \sin^2\theta}\right)
    \eqbreakadd - \frac{1}{4\pi\epsilon_0} \frac{e^2}{r}.
\end{align*}
The action of $S_T$ on $H'$ results in the following function
\begin{align*}
& (S_T^{-1}H')(r,\theta,\phi,p_r,p_\theta,p_\phi) = \frac{1}{2m}\left(p_r^2
    + \frac{p_\theta^2}{r^2} + \frac{p_\phi^2}{r^2 \sin^2\theta}\right)
    \eqbreakadd - \frac{1}{4\pi\epsilon_0} \frac{e^2}{r}
- \frac{\hbar^2}{8mr^2}\left(\frac{1}{\sin^2\theta} + 1\right).
\end{align*}
From this to $H'$ we can associate the following operator being a symmetrically
ordered function $S_T^{-1}H'$ of operators of position
$\hat{q}_r,\hat{q}_\theta,\hat{q}_\phi$ and momentum
$\hat{p}_r,\hat{p}_\theta,\hat{p}_\phi$:
\begin{equation*}
H'_{S_T}(\hat{q}_r,\hat{q}_\theta,\hat{q}_\phi,\hat{p}_r,\hat{p}_\theta,
    \hat{p}_\phi) = -\frac{\hbar^2}{2m}\left[\partial_r^2
    + \frac{2}{r}\partial_r + \frac{1}{r^2}\left(\partial_\theta^2
    + \frac{1}{\tan\theta} \partial_\theta + \frac{1}{\sin^2\theta}
    \partial_\phi^2 \right)\right] - \frac{1}{4\pi\epsilon_0} \frac{e^2}{r}.
\end{equation*}
Note that the expression in square brackets is just the Laplace operator
written in spherical coordinates.
\end{example}

\begin{remark}
It has to be stressed that for a particular class of point transformations
$T$ \eqref{eq:4.16}, defined on almost the whole phase space and taking values
in almost the whole phase space, i.e. $T \colon \mathbb{R}^{2N} \supset V \times
\mathbb{R}^N \to U \times \mathbb{R}^N \subset \mathbb{R}^{2N}$, where
$\mathbb{R}^N \setminus U$ and $\mathbb{R}^N \setminus V$ are sets of the
Lebesgue-measure zero, there exists an alternative operator representation of
observables in a Hilbert space $L^2(V)$ with Lebesgue-measure in new
coordinates, instead in a Hilbert space $L^2(V,\mu)$. For that representation
operators of position and momentum in new coordinates have the same form as in
the pseudo-Euclidean case, the form of $S_T$ operator remains the same but the
unitary operator $\hat{U}_T$ \eqref{eq:4.15} takes a different form. What is
important is that both representations describe the same quantum systems. Such a
construction can be extended onto a wider class of canonical transformations
than the point transformations. The details of the construction the reader can
find in \cite{Blaszak:2013}.
\end{remark}

\subsection{Invariant representation of Hamilton operators}
\label{subsec:4.3}
Until now we were considering the Hilbert space approach to quantum mechanics
in a representation corresponding to some coordinate system on the configuration
space. It is, however, possible to consider the Hilbert space approach to
quantum mechanics in a coordinate independent way. In such an approach the
Hilbert space of states is taken to be the space $L^2(\mathcal{Q},\omega_g)$ of
square integrable functions defined on the configuration space $\mathcal{Q}$
with respect to the metric volume form $\omega_g$. If we choose some coordinate
system on $\mathcal{Q}$: $\phi \colon U \subset \mathcal{Q} \to V \subset
\mathbb{R}^N$, $\phi(P) = (x^1,\dotsc,x^N)$ then we can define observables of
position $\hat{q}^i$ for this coordinate system as multiplication operators by
$\phi^i$:
\begin{equation*}
(\hat{q}^i\psi)(P) = \phi^i(P)\psi(P).
\end{equation*}
The operators $\hat{q}^1,\dotsc,\hat{q}^N$ constitute the complete set of
commuting observables and can be used to create the representation corresponding
to the coordinate system $\phi$. In this representation operators $\hat{q}^i$
take the form of the multiplication operators by a coordinate variable, and the
Hilbert space of states takes the form of the space $L^2(V,\mu)$, where
$\dd{\mu(x)} = \abs{\det[g_{ij}(x)]}^{1/2} \ud{x}$.

Using the previous results we will show how to write Hamiltonians quadratic and
cubic in momenta in an invariant way. First let us consider a Hamiltonian $H$
quadratic in momenta, which in a pseudo-Euclidean coordinate system takes the
form
\begin{equation*}
H(x,p) = \frac{1}{2} K^{ij}(x) p_i p_j + V(x),
\end{equation*}
where $K^{ij}$ are components of some symmetric tensor $K$. After performing a
point transformation \eqref{eq:2.13} the Hamiltonian $H$ can be
written in the form
\begin{equation}
H'(x',p') = \frac{1}{2} K'^{ij}(x') p'_i p'_j + V(x'),
\label{eq:4.13}
\end{equation}
where $K'^{ij}(x')$ are components of the tensor $K$ for $(x'^1,\dotsc,x'^N)$
coordinates.

The action of $S_T$ \eqref{eq:2.17} on $H'$ results in the following
function
\begin{equation*}
(S_T^{-1} H')(x',p') = \frac{1}{2} K'^{ij}(x') p'_i p'_j + V(x')
-\frac{\hbar^2}{2}\left(\frac{1}{4}K'^{ij}_{\phantom{ij},k}(x')\Gamma^k_{ij}(x')
+\frac{1}{4}K'^{ij}(x')\Gamma^k_{li}(x')\Gamma^l_{kj}(x') \right),
\end{equation*}
where $,k$ denotes the partial derivative with respect to $x'^k$.
From this to $H'$ will correspond the following operator
\begin{align*}
H'_{S_T}(\hat{q}',\hat{p}') & = \frac{1}{2} \left( \frac{1}{4}K'^{ij}(\hat{q}')
    \hat{p}'_i \hat{p}'_j + \frac{1}{2} \hat{p}'_i K'^{ij}(\hat{q}') \hat{p}'_j
+ \frac{1}{4} \hat{p}'_i \hat{p}'_j K'^{ij}(\hat{q}') \right) + V(\hat{q}')
\nonumber \\
& \quad {} -\frac{\hbar^2}{2}\left(\frac{1}{4}K'^{ij}_{\phantom{ij},k}(\hat{q}')
    \Gamma^k_{ij}(\hat{q}')
+ \frac{1}{4}K'^{ij}(\hat{q}')\Gamma^k_{li}(\hat{q}')\Gamma^l_{kj}(\hat{q}')
    \right).
\end{align*}
By virtue of \eqref{eq:4.8} the above equation can be written in the form
\begin{align*}
H'_{S_T}(\hat{q}',\hat{p}') & = -\frac{\hbar^2}{2} \biggl(
    K'^{ij} \partial_{x'^i}\partial_{x'^j} + K'^{ij}\Gamma^l_{jl}\partial_{x'^i}
+ K'^{ij}_{\phantom{ij},i}\partial_{x'^j} + \frac{1}{2}K'^{ij}\Gamma^l_{jl,i}
+ \frac{1}{4}K'^{ij}\Gamma^k_{ik}\Gamma^l_{jl}
+ \frac{1}{2}K'^{ij}_{\phantom{ij},i}\Gamma^l_{jl} \nonumber \\
& \quad {} + \frac{1}{4}K'^{ij}_{\phantom{ij},ij}
+ \frac{1}{4}K'^{ij}_{\phantom{ij},k}\Gamma^k_{ij}
+ \frac{1}{4}K'^{ij}\Gamma^k_{li}\Gamma^l_{kj} \biggr) + V.
\end{align*}
Using the equality $K'^{ij}_{\phantom{ij},k} = -K'^{rj}\Gamma^i_{rk} -
K'^{ri}\Gamma^j_{rk} + K'^{ij}_{\phantom{ij};k}$ where $;k$ denotes the
covariant derivative in the direction of the vector field $\partial_{x'^k}$,
and the flatness of the connection $\nabla$ on the configuration space, the
above equation simplifies to
\begin{align}
H'_{S_T}(\hat{q}',\hat{p}') & = -\frac{\hbar^2}{2} \biggl(
    K'^{ij} \partial_{x'^i}\partial_{x'^j} + K'^{ij}\Gamma^l_{jl}\partial_{x'^i}
    \eqbreakadd + K'^{ij}_{\phantom{ij},i}\partial_{x'^j}
    + \frac{1}{4} K'^{ij}_{\phantom{ij};ij} \biggr) + V.
\label{eq:4.9}
\end{align}
Note, that \eqref{eq:4.9} can be written in the following form
\begin{equation}
H'_{S_T}(\hat{q}',\hat{p}') = -\frac{\hbar^2}{2} \left(\nabla_i K'^{ij} \nabla_j
+ \frac{1}{4} K'^{ij}_{\phantom{ij};ij}\right) + V,
\label{eq:4.10}
\end{equation}
where $\nabla_i K'^{ij} \nabla_j = \Delta_K$ is the pseudo-Laplace operator.
For a special case when $K$ is the standard metric tensor $g$ on the
configuration space, the Hamiltonian $H$ has the form of a natural Hamiltonian
and \eqref{eq:4.10} reduces to
\begin{equation*}
H'_{S_T}(\hat{q}',\hat{p}') = -\frac{\hbar^2}{2}g'^{ij} \nabla_i \nabla_j + V.
\label{eq:4.11}
\end{equation*}
Observe, that $\nabla_i g'^{ij} \nabla_j = g'^{ij} \nabla_i \nabla_j = \Delta$
is the Laplace operator in curvilinear coordinates.

Let us now consider a Hamiltonian $H$, which in a pseudo-Euclidean coordinate
system is cubic in momenta (we skip the lower terms in momenta):
\begin{equation*}
H(x,p) = K^{ijk}(x) p_i p_j p_k,
\end{equation*}
where $K^{ijk}$ are components of some symmetric tensor $K$. In $(x',p')$
coordinates the Hamiltonian $H$ can be written in the form
\begin{equation}
H'(x',p') = K'^{ijk}(x') p'_i p'_j p'_k,
\label{eq:4.14}
\end{equation}
where $K'^{ijk}(x')$ are components of the tensor $K$ for $(x'^1,\dotsc,x'^N)$
coordinates.

The action of $S_T$ on $H'$ results in the following function
\begin{align*}
(S_T^{-1} H')(x',p') & = K'^{ijk}(x') p'_i p'_j p'_k
- \frac{\hbar^2}{4} \Bigl(3\Gamma^i_{jk}(x') K'^{ljk}_{\phantom{ljk},i}(x') p'_l
+ 3\Gamma^i_{lj}(x') \Gamma^l_{ik}(x') K'^{rjk}(x') p'_r \nonumber \\
& \quad {} + \left(2\Gamma^i_{rl}(x') \Gamma^r_{jk}(x')
- \Gamma^i_{jk,l}(x')\right) K'^{jkl}(x') p'_i \Bigr).
\end{align*}
From this to $H'$ will correspond the following operator
\begin{align}
H'_{S_T}(\hat{q}',\hat{p}') & = i\hbar^3 \biggl(
    K'^{ijk} \partial_{x'^i} \partial_{x'^j} \partial_{x'^k}
    + \frac{3}{2} K'^{ijk}_{\phantom{ijk},i} \partial_{x'^j} \partial_{x'^k}
    - 3 K'^{ijk} \Gamma^l_{ij} \partial_{x'^l} \partial_{x'^k}
    + \frac{3}{4} K'^{ijk}_{\phantom{ijk};ij} \partial_{x'^k} \nonumber \\
& \quad {} - \frac{3}{2} K'^{ijk}_{\phantom{ijk};i} \Gamma^l_{jk}\partial_{x'^l}
    + 2 K'^{ijk} \Gamma^l_{rk} \Gamma^r_{ij} \partial_{x'^l}
    - K'^{ijk} \Gamma^l_{ij,k} \partial_{x'^l}
    + \frac{1}{8} K'^{ijk}_{\phantom{ijk};ijk} \biggr) \nonumber \\
& = \frac{1}{2}i\hbar^3 \biggl( \nabla_i K'^{ijk} \nabla_j \nabla_k
    + \nabla_i \nabla_j K'^{ijk} \nabla_k
    + \frac{1}{2} K'^{ijk}_{\phantom{ijk};ij} \nabla_k
    + \frac{1}{4} K'^{ijk}_{\phantom{ijk};ijk} \biggr) \nonumber \\
& = \frac{1}{2}i\hbar^3 \biggl( \nabla_i K'^{ijk} \nabla_j \nabla_k
    + \nabla_i \nabla_j K'^{ijk} \nabla_k
    + \frac{1}{4} \nabla_k K'^{ijk}_{\phantom{ijk};ij}
    + \frac{1}{4} K'^{ijk}_{\phantom{ijk};ij} \nabla_k \biggr).
\label{eq:4.12}
\end{align}

Note that we received operators \eqref{eq:4.10} and \eqref{eq:4.12} written in
a coordinate independent way. Although these operators are defined on a Hilbert
space $L^2(V,\mu)$ corresponding to a particular coordinate system, we can treat
these operators as defined on a Hilbert space $L^2(\mathcal{Q},\omega_g)$.

\section{Remarks on quantization in curved spaces}
\label{sec:5}

\subsection{Admissible invariant quantum Hamiltonians}
\label{subsec:5.1}
Until now we were considering quantization of classical systems over flat
configuration spaces. In the following section we will discus how to quantize
systems over curved configuration spaces. Let us take as the configuration space
$\mathcal{Q}$ the Riemannian manifold $(\mathbb{R}^N,g)$, where $g$ is some
general non-flat metric tensor of signature $(r,s)$. To quantize a classical
system defined over the configuration space $\mathcal{Q}$ it is necessary to
introduce a star-product over the phase space $M = T^*\mathcal{Q}$. This
product, after writing it in some quantum canonical coordinate system, should be
equivalent with the Moyal product in the sense of \subsecref{subsec:2.3}, and
for a flat case and coordinate system induced from a coordinate system on the
configuration space it should be of the form \eqref{eq:2.2}. The simplest way of
receiving such star-product is by defining, for some coordinate system
$(x^1,\dotsc,x^N)$ on the configuration space, an isomorphism $S$ which would
reduce, for a flat connection on $\mathcal{Q}$, to an isomorphism given by
\eqref{eq:2.17}. Then the isomorphism $S$ can be used to define an admissible
star-product by acting on a Moyal product. Of course there exist infinitely many
such isomorphisms $S$ and related quantizations. Which quantizations are
``proper'' could only be verified by some additional physical arguments, if one
could find them. Let us present the following family of quantizations defined by
the following family of isomorphisms $S$:
\begin{align}
S & = \id + \frac{\hbar^2}{4!} \Bigl(3\left(\Gamma^i_{lj}(x)\Gamma^l_{ik}(x)
    + \alpha R_{jk}(x)\right)\partial_{p_j}\partial_{p_k} \eqbreakadd
    + 3\Gamma^i_{jk}(x)\partial_{x^i}\partial_{p_j}\partial_{p_k}
    \nonumber \\
& \quad {} + \left( 2\Gamma^i_{nl}(x)\Gamma^n_{jk}(x)
    - \partial_{x^l}\Gamma^i_{jk}(x) \right)
    p_i \partial_{p_j}\partial_{p_k}\partial_{p_l} \Bigr) \eqbreakadd
    + o(\hbar^4),
\label{eq:5.1}
\end{align}
where $R_{jk}$ is the Ricci curvature tensor and $\alpha \in \mathbb{R}$. Of
course in a flat case $R_{jk} = 0$ and \eqref{eq:5.1} reduces to
\eqref{eq:2.17}.

The passage to the operator representation over the configuration space of
quantum mechanics can be made in a similar fashion as in \subsecref{subsec:4.2}.
For some coordinate system $(x^1,\dotsc,x^N)$ on the configuration space we can
define the Hilbert space of states as $L^2(\mathbb{R}^N,\mu)$, where
$\dd{\mu(x)} = \abs{\det[g_{ij}(x)]}^{1/2} \ud{x}$, and operators of position
and momentum as
\begin{subequations}
\begin{align*}
\hat{q}^i & = x^i, \\
\hat{p}_j & = -i\hbar\left(\partial_{x^j} + \frac{1}{2}\Gamma^k_{jk}(x)\right).
\end{align*}
\end{subequations}
Using \eqref{eq:5.1} and performing similar calculations as in
\subsecref{subsec:4.3} we can derive the expressions for operators associated
with Hamilton functions quadratic and cubic in momenta defined on a curved
space. For a Hamiltonian quadratic in momenta we receive
\begin{align*}
H_S(\hat{q},\hat{p}) & = -\frac{\hbar^2}{2} \biggl(\nabla_i K^{ij} \nabla_j
    + \frac{1}{4} K^{ij}_{\phantom{ij};ij} \eqbreakadd
    - \frac{1}{4}(1 - \alpha) K^{ij} R_{ij} \biggr) + V.
\end{align*}
When $K^{ij}$ is the metric tensor $g^{ij}$ the above formula reduces to
\begin{equation}
H_S(\hat{q},\hat{p}) = -\frac{\hbar^2}{2} \left(g^{ij}\nabla_i \nabla_j
- \frac{1}{4}(1 - \alpha) R \right) + V,
\label{eq:5.2}
\end{equation}
where $R$ is the scalar curvature. Note that \eqref{eq:5.2} for particular
values of the parameter $\alpha$ is the form of the Hamiltonian operator
quadratic in momenta derived by the use of various techniques
\cite{DeWitt:1957,Dekker:1980,Liu:1992,Duval:2001}. For a Hamiltonian cubic
in momenta we receive
\begin{align*}
H_S(\hat{q},\hat{p}) & = \frac{1}{2} i\hbar^3 \biggl(
    \nabla_i K^{ijk} \nabla_j \nabla_k + \nabla_i \nabla_j K^{ijk} \nabla_k
    + \frac{1}{4} \nabla_k K^{ijk}_{\phantom{ijk};ij}
    + \frac{1}{4} K^{ijk}_{\phantom{ijk};ij} \nabla_k \nonumber \\
& \quad {} - \frac{3}{4}(1 - \alpha) \nabla_i K^{ijk} R_{jk}
    - \frac{3}{4}(1 - \alpha) K^{ijk} R_{jk} \nabla_i \biggr).
\end{align*}

\subsection{On ambiguity of quantization}
\label{subsec:5.2}
In previous sections we developed an invariant quantization theory based on the
canonical choice from \subsecref{subsec:2.3}, i.e. using a Moyal star product in
pseudo-Euclidean coordinates and quantum observables equal to classical ones.
Here we analyze a different admissible choice. Let us consider another family of
invariant star product related to the decomposition \eqref{eq:2.1} of the
classical Poisson tensor $\mathcal{P}$
\begin{align}
f \star g & = f \exp \Biggl(
    \frac{1}{2} i\hbar \sum_k\overleftarrow{X}_k \overrightarrow{Y}_k
    - \frac{1}{2} i\hbar \sum_k\overleftarrow{Y}_k \overrightarrow{X}_k
    + P(\overleftarrow{X}_1 + \overrightarrow{X}_1, \dotsc, \overleftarrow{Y}_N
    + \overrightarrow{Y}_N; \hbar) \nonumber \\
& \quad {} - P(\overleftarrow{X}_1,\dotsc,\overleftarrow{Y}_N;\hbar)
    - P(\overrightarrow{X}_1,\dotsc,\overrightarrow{Y}_N;\hbar) \Biggr) g,
\label{eq:5.5}
\end{align}
where $P$ is some polynomial of $2N$ variables with coefficients dependent
on $\hbar$, such that
\begin{equation*}
\overline{P(X_1,\dotsc,Y_N)} = P(Y_1,\dotsc,X_N).
\end{equation*}
What is important, the complex-conjugation is the involution for this product as
well. An isomorphism \eqref{eq:2.5} intertwining the \nbr{\star}product
\eqref{eq:5.5} with the \nbr{\star}product \eqref{eq:2.2} reads
\begin{equation*}
S = \exp\left(P(X_1,\dotsc,Y_N;\hbar)\right).
\label{eq:5.6}
\end{equation*}
As an example let us take $P(X_1,\dotsc,Y_N;\hbar) = -\frac{1}{8} \hbar^2
\sum_{k,j}X_k X_j Y_k Y_j$. Then the \nbr{\star}product \eqref{eq:5.5} takes the
form
\begin{align}
f \star g & = f \exp \Biggl(
    \frac{1}{2} i\hbar \sum_k\overleftarrow{X}_k \overrightarrow{Y}_k
    - \frac{1}{2} i\hbar \sum_k\overleftarrow{Y}_k \overrightarrow{X}_k
    + \frac{1}{8} \hbar^2 \sum_{k,j} (\overleftarrow{X}_k \overleftarrow{Y}_k
    \overleftarrow{X}_j \overleftarrow{Y}_j + \overrightarrow{X}_k
    \overrightarrow{Y}_k \overrightarrow{X}_j \overrightarrow{Y}_j) \nonumber \\
& \quad {} - \frac{1}{8} \hbar^2 \sum_{k,j}
    (\overleftarrow{X}_k + \overrightarrow{X}_k)
    (\overleftarrow{Y}_k + \overrightarrow{Y}_k)
    (\overleftarrow{X}_j + \overrightarrow{X}_j)
    (\overleftarrow{Y}_j + \overrightarrow{Y}_j) \Biggr) g.
\label{eq:5.10}
\end{align}

Now, let us choose as the canonical \nbr{\star}product in a flat case the
product \eqref{eq:5.10} with $X_i = \partial_{x^i}$, $Y_i = \partial_{p_i}$ in a
pseudo-Euclidean coordinates $(x,p)$ and choose the quantum observables $A_Q$
equal exactly to the classical ones $A_C$. Such a quantization is equivalent
with the choice of standard Moyal \nbr{\star}product with another choice of
quantum observables. Actually, for any curvilinear coordinates
\begin{equation}
A_Q = \exp\left(\frac{1}{8} \hbar^2 \sum_{k,j}\nabla_k \nabla_j \partial_{p_k}
    \partial_{p_j}\right) A_C.
\label{eq:5.7}
\end{equation}
Now, invariant quantization of a quadratic in momenta classical Hamiltonian
\eqref{eq:4.13} gives the operator
\begin{equation}
(H_Q)_{S_T}(\hat{q},\hat{p}) = -\frac{\hbar^2}{2} \nabla_i K^{ij} \nabla_j + V,
\label{eq:5.8}
\end{equation}
and for cubic in momenta term \eqref{eq:4.14} the related operator form
\begin{equation}
(H_Q)_{S_T}(\hat{q},\hat{p}) = \frac{1}{2}i\hbar^3 \biggl(
    \nabla_i K^{ijk} \nabla_j \nabla_k
    + \nabla_i \nabla_j K^{ijk} \nabla_k \biggr).
\label{eq:5.9}
\end{equation}
The extension onto a non-flat case remains the same except the new form of
quantum observable \eqref{eq:5.7}. So, with the particular choice $\alpha = 1$
in $S$ \eqref{eq:5.1}, operators \eqref{eq:5.8} and \eqref{eq:5.9} are
admissible quantum Hamiltonians for classical systems quadratic and cubic in
momenta in any Riemann space. Such choice of quantization was called in a paper
\cite{Duval:2005} a ``minimal'' quantization, but was introduced \emph{ad hoc}
without any justification from basic principles. Moreover, the same choice was
made in \cite{Benenti:2002a,Benenti:2002b} in order to investigate the quantum
integrability and quantum separability of classical St\"ackel systems.

%


\begin{thebibliography}{26}%
\makeatletter
\providecommand \@ifxundefined [1]{%
 \@ifx{#1\undefined}
}%
\providecommand \@ifnum [1]{%
 \ifnum #1\expandafter \@firstoftwo
 \else \expandafter \@secondoftwo
 \fi
}%
\providecommand \@ifx [1]{%
 \ifx #1\expandafter \@firstoftwo
 \else \expandafter \@secondoftwo
 \fi
}%
\providecommand \natexlab [1]{#1}%
\providecommand \enquote  [1]{``#1''}%
\providecommand \bibnamefont  [1]{#1}%
\providecommand \bibfnamefont [1]{#1}%
\providecommand \citenamefont [1]{#1}%
\providecommand \href@noop [0]{\@secondoftwo}%
\providecommand \href [0]{\begingroup \@sanitize@url \@href}%
\providecommand \@href[1]{\@@startlink{#1}\@@href}%
\providecommand \@@href[1]{\endgroup#1\@@endlink}%
\providecommand \@sanitize@url [0]{\catcode `\\12\catcode `\$12\catcode
  `\&12\catcode `\#12\catcode `\^12\catcode `\_12\catcode `\%12\relax}%
\providecommand \@@startlink[1]{}%
\providecommand \@@endlink[0]{}%
\providecommand \url  [0]{\begingroup\@sanitize@url \@url }%
\providecommand \@url [1]{\endgroup\@href {#1}{\urlprefix }}%
\providecommand \urlprefix  [0]{URL }%
\providecommand \Eprint [0]{\href }%
\providecommand \doibase [0]{http://dx.doi.org/}%
\providecommand \selectlanguage [0]{\@gobble}%
\providecommand \bibinfo  [0]{\@secondoftwo}%
\providecommand \bibfield  [0]{\@secondoftwo}%
\providecommand \translation [1]{[#1]}%
\providecommand \BibitemOpen [0]{}%
\providecommand \bibitemStop [0]{}%
\providecommand \bibitemNoStop [0]{.\EOS\space}%
\providecommand \EOS [0]{\spacefactor3000\relax}%
\providecommand \BibitemShut  [1]{\csname bibitem#1\endcsname}%
\let\auto@bib@innerbib\@empty
\bibitem [{\citenamefont {Podolsky}(1928)}]{Podolsky:1928}%
  \BibitemOpen
  \bibfield  {author} {\bibinfo {author} {\bibfnamefont {B.}~\bibnamefont
  {Podolsky}},\ }\href@noop {} {\bibfield  {journal} {\bibinfo  {journal}
  {Phys. Rev.}\ }\textbf {\bibinfo {volume} {32}},\ \bibinfo {pages} {812}
  (\bibinfo {year} {1928})}\BibitemShut {NoStop}%
\bibitem [{\citenamefont {DeWitt}(1952)}]{DeWitt:1952}%
  \BibitemOpen
  \bibfield  {author} {\bibinfo {author} {\bibfnamefont {B.~S.}\ \bibnamefont
  {DeWitt}},\ }\href@noop {} {\bibfield  {journal} {\bibinfo  {journal} {Phys.
  Rev.}\ }\textbf {\bibinfo {volume} {85}},\ \bibinfo {pages} {653} (\bibinfo
  {year} {1952})}\BibitemShut {NoStop}%
\bibitem [{\citenamefont {DeWitt}(1957)}]{DeWitt:1957}%
  \BibitemOpen
  \bibfield  {author} {\bibinfo {author} {\bibfnamefont {B.~S.}\ \bibnamefont
  {DeWitt}},\ }\href@noop {} {\bibfield  {journal} {\bibinfo  {journal} {Rev.
  Mod. Phys.}\ }\textbf {\bibinfo {volume} {29}},\ \bibinfo {pages} {377}
  (\bibinfo {year} {1957})}\BibitemShut {NoStop}%
\bibitem [{\citenamefont {Gervais}\ and\ \citenamefont
  {Jevicki}(1976)}]{Gervais:1976}%
  \BibitemOpen
  \bibfield  {author} {\bibinfo {author} {\bibfnamefont {J.-L.}\ \bibnamefont
  {Gervais}}\ and\ \bibinfo {author} {\bibfnamefont {A.}~\bibnamefont
  {Jevicki}},\ }\href@noop {} {\bibfield  {journal} {\bibinfo  {journal} {Nucl.
  Phys. B}\ }\textbf {\bibinfo {volume} {110}},\ \bibinfo {pages} {93} (\bibinfo
  {year} {1976})}\BibitemShut {NoStop}%
\bibitem [{\citenamefont {Carter}(1977)}]{Carter:1977}%
  \BibitemOpen
  \bibfield  {author} {\bibinfo {author} {\bibfnamefont {B.}~\bibnamefont
  {Carter}},\ }\href@noop {} {\bibfield  {journal} {\bibinfo  {journal} {Phys.
  Rev. D}\ }\textbf {\bibinfo {volume} {16}},\ \bibinfo {pages} {3395}
  (\bibinfo {year} {1977})}\BibitemShut {NoStop}%
\bibitem [{\citenamefont {Ess{\'e}n}(1978)}]{Essen:1978}%
  \BibitemOpen
  \bibfield  {author} {\bibinfo {author} {\bibfnamefont {H.}~\bibnamefont
  {Ess{\'e}n}},\ }\href@noop {} {\bibfield  {journal} {\bibinfo  {journal} {Am.
  J. Phys.}\ }\textbf {\bibinfo {volume} {46}},\ \bibinfo {pages} {983}
  (\bibinfo {year} {1978})}\BibitemShut {NoStop}%
\bibitem [{\citenamefont {Dekker}(1980)}]{Dekker:1980}%
  \BibitemOpen
  \bibfield  {author} {\bibinfo {author} {\bibfnamefont {H.}~\bibnamefont
  {Dekker}},\ }\href@noop {} {\bibfield  {journal} {\bibinfo  {journal}
  {Physica A (Utrecht)}\ }\textbf {\bibinfo {volume} {103}},\ \bibinfo {pages}
  {586} (\bibinfo {year} {1980})}\BibitemShut {NoStop}%
\bibitem [{\citenamefont {Liu}\ and\ \citenamefont {Quian}(1992)}]{Liu:1992}%
  \BibitemOpen
  \bibfield  {author} {\bibinfo {author} {\bibfnamefont {Z.~J.}\ \bibnamefont
  {Liu}}\ and\ \bibinfo {author} {\bibfnamefont {M.}~\bibnamefont {Quian}},\
  }\href@noop {} {\bibfield  {journal} {\bibinfo  {journal} {Trans. Amer. Math.
  Soc.}\ }\textbf {\bibinfo {volume} {331}},\ \bibinfo {pages} {321} (\bibinfo
  {year} {1992})}\BibitemShut {NoStop}%
\bibitem [{\citenamefont {Duval}\ and\ \citenamefont
  {Ovsienko}(2001)}]{Duval:2001}%
  \BibitemOpen
  \bibfield  {author} {\bibinfo {author} {\bibfnamefont {C.}~\bibnamefont
  {Duval}}\ and\ \bibinfo {author} {\bibfnamefont {V.}~\bibnamefont
  {Ovsienko}},\ }\href@noop {} {\bibfield  {journal} {\bibinfo  {journal}
  {Selecta Math. New Ser.}\ }\textbf {\bibinfo {volume} {7}},\ \bibinfo {pages}
  {291} (\bibinfo {year} {2001})}\BibitemShut {NoStop}%
\bibitem [{\citenamefont {Loubon~Djounga}(2003)}]{LoubonDjounga:2003}%
  \BibitemOpen
  \bibfield  {author} {\bibinfo {author} {\bibfnamefont {S.~E.}\ \bibnamefont
  {Loubon~Djounga}},\ }\href@noop {} {\bibfield  {journal} {\bibinfo  {journal}
  {Lett. Math. Phys.}\ }\textbf {\bibinfo {volume} {64}},\ \bibinfo {pages}
  {203} (\bibinfo {year} {2003})}\BibitemShut {NoStop}%
\bibitem [{\citenamefont {Duval}\ and\ \citenamefont
  {Valent}(2005)}]{Duval:2005}%
  \BibitemOpen
  \bibfield  {author} {\bibinfo {author} {\bibfnamefont {C.}~\bibnamefont
  {Duval}}\ and\ \bibinfo {author} {\bibfnamefont {G.}~\bibnamefont {Valent}},\
  }\href@noop {} {\bibfield  {journal} {\bibinfo  {journal} {J. Math. Phys.}\
  }\textbf {\bibinfo {volume} {46}},\ \bibinfo {pages} {053516} (\bibinfo
  {year} {2005})}\BibitemShut {NoStop}%
\bibitem [{\citenamefont {Moyal}(1949)}]{Moyal:1949}%
  \BibitemOpen
  \bibfield  {author} {\bibinfo {author} {\bibfnamefont {J.~E.}\ \bibnamefont
  {Moyal}},\ }\href@noop {} {\bibfield  {journal} {\bibinfo  {journal} {Proc.
  Cambridge Philos. Soc.}\ }\textbf {\bibinfo {volume} {45}},\ \bibinfo {pages}
  {99} (\bibinfo {year} {1949})}\BibitemShut {NoStop}%
\bibitem [{\citenamefont {Bayen}\ \emph
  {et~al.}(1978{\natexlab{a}})\citenamefont {Bayen}, \citenamefont {Flato},
  \citenamefont {Fr{\o}nsdal}, \citenamefont {Lichnerowicz},\ and\
  \citenamefont {Sternheimer}}]{Bayen:1978a}%
  \BibitemOpen
  \bibfield  {author} {\bibinfo {author} {\bibfnamefont {F.}~\bibnamefont
  {Bayen}}, \bibinfo {author} {\bibfnamefont {M.}~\bibnamefont {Flato}},
  \bibinfo {author} {\bibfnamefont {C.}~\bibnamefont {Fr{\o}nsdal}}, \bibinfo
  {author} {\bibfnamefont {A.}~\bibnamefont {Lichnerowicz}}, \ and\ \bibinfo
  {author} {\bibfnamefont {D.}~\bibnamefont {Sternheimer}},\ }\href@noop {}
  {\bibfield  {journal} {\bibinfo  {journal} {Ann. Phys.}\ }\textbf {\bibinfo
  {volume} {111}},\ \bibinfo {pages} {61} (\bibinfo {year}
  {1978}{\natexlab{a}})}\BibitemShut {NoStop}%
\bibitem [{\citenamefont {Bayen}\ \emph
  {et~al.}(1978{\natexlab{b}})\citenamefont {Bayen}, \citenamefont {Flato},
  \citenamefont {Fr{\o}nsdal}, \citenamefont {Lichnerowicz},\ and\
  \citenamefont {Sternheimer}}]{Bayen:1978b}%
  \BibitemOpen
  \bibfield  {author} {\bibinfo {author} {\bibfnamefont {F.}~\bibnamefont
  {Bayen}}, \bibinfo {author} {\bibfnamefont {M.}~\bibnamefont {Flato}},
  \bibinfo {author} {\bibfnamefont {C.}~\bibnamefont {Fr{\o}nsdal}}, \bibinfo
  {author} {\bibfnamefont {A.}~\bibnamefont {Lichnerowicz}}, \ and\ \bibinfo
  {author} {\bibfnamefont {D.}~\bibnamefont {Sternheimer}},\ }\href@noop {}
  {\bibfield  {journal} {\bibinfo  {journal} {Ann. Phys.}\ }\textbf {\bibinfo
  {volume} {111}},\ \bibinfo {pages} {111} (\bibinfo {year}
  {1978}{\natexlab{b}})}\BibitemShut {NoStop}%
\bibitem [{\citenamefont {Dito}\ and\ \citenamefont
  {Sternheimer}(2002)}]{Dito.Sternheimer:2002}%
  \BibitemOpen
  \bibfield  {author} {\bibinfo {author} {\bibfnamefont {G.}~\bibnamefont
  {Dito}}\ and\ \bibinfo {author} {\bibfnamefont {D.}~\bibnamefont
  {Sternheimer}},\ }in\ \href@noop {} {\emph {\bibinfo {booktitle} {Deformation
  quantization}}},\ \bibinfo {series} {{IRMA} lectures in mathematics and
  theoretical physics}, Vol.~\bibinfo {volume} {1},\ \bibinfo {editor} {edited
  by\ \bibinfo {editor} {\bibfnamefont {G.}~\bibnamefont {Halbout}}}\ (\bibinfo
   {publisher} {Walter de Gruyter},\ \bibinfo {address} {Berlin, New York},\
  \bibinfo {year} {2002})\ pp.\ \bibinfo {pages} {9--54}\BibitemShut {NoStop}%
\bibitem [{\citenamefont {Gutt}(2000)}]{Gutt:2000}%
  \BibitemOpen
  \bibfield  {author} {\bibinfo {author} {\bibfnamefont {S.}~\bibnamefont
  {Gutt}},\ }in\ \href@noop {} {\emph {\bibinfo {booktitle} {Conf{\'e}rence
  {M}osh{\'e} {F}lato 1999: quantization, deformations, and symmetries}}},\
  \bibinfo {series} {Mathematical physics studies}, Vol.~\bibinfo {volume}
  {21},\ \bibinfo {editor} {edited by\ \bibinfo {editor} {\bibfnamefont
  {G.}~\bibnamefont {Dito}}\ and\ \bibinfo {editor} {\bibfnamefont
  {D.}~\bibnamefont {Sternheimer}}}\ (\bibinfo  {publisher} {Kluwer Academic
  Publishers},\ \bibinfo {address} {Netherlands},\ \bibinfo {year} {2000})\
  pp.\ \bibinfo {pages} {217--254}\BibitemShut {NoStop}%
\bibitem [{\citenamefont {Weinstein}(1994)}]{Weinstein:1994b}%
  \BibitemOpen
  \bibfield  {author} {\bibinfo {author} {\bibfnamefont {A.}~\bibnamefont
  {Weinstein}},\ }in\ \href@noop {} {\emph {\bibinfo {booktitle} {S{\'e}minaire
  Bourbaki}}},\ Vol.~\bibinfo {volume} {36}\ (\bibinfo  {publisher}
  {Association des Collaborateurs de Nicolas Bourbaki},\ \bibinfo {year}
  {1993--1994})\ pp.\ \bibinfo {pages} {389--409}\BibitemShut {NoStop}%
\bibitem [{\citenamefont {Curtright}\ \emph {et~al.}(1998)\citenamefont
  {Curtright}, \citenamefont {Fairlie},\ and\ \citenamefont
  {Zachos}}]{Curtright:1998}%
  \BibitemOpen
  \bibfield  {author} {\bibinfo {author} {\bibfnamefont {T.}~\bibnamefont
  {Curtright}}, \bibinfo {author} {\bibfnamefont {D.~B.}\ \bibnamefont
  {Fairlie}}, \ and\ \bibinfo {author} {\bibfnamefont {C.}~\bibnamefont
  {Zachos}},\ }\href@noop {} {\bibfield  {journal} {\bibinfo  {journal} {Phys.
  Rev. D}\ }\textbf {\bibinfo {volume} {58}},\ \bibinfo {pages} {25002}
  (\bibinfo {year} {1998})},\ \Eprint
  {http://arxiv.org/abs/arXiv:hep-th/9711183v3} {arXiv:hep-th/9711183v3}
  \BibitemShut {NoStop}%
\bibitem [{\citenamefont {Curtright}\ and\ \citenamefont
  {Zachos}(1999)}]{Curtright:1999a}%
  \BibitemOpen
  \bibfield  {author} {\bibinfo {author} {\bibfnamefont {T.}~\bibnamefont
  {Curtright}}\ and\ \bibinfo {author} {\bibfnamefont {C.}~\bibnamefont
  {Zachos}},\ }\href@noop {} {\bibfield  {journal} {\bibinfo  {journal} {J.
  Phys. A}\ }\textbf {\bibinfo {volume} {32}},\ \bibinfo {pages} {771}
  (\bibinfo {year} {1999})},\ \Eprint
  {http://arxiv.org/abs/arXiv:hep-th/9810164v2} {arXiv:hep-th/9810164v2}
  \BibitemShut {NoStop}%
\bibitem [{\citenamefont {B{\l}aszak}\ and\ \citenamefont
  {Doma{\'n}ski}(2012)}]{Blaszak:2012}%
  \BibitemOpen
  \bibfield  {author} {\bibinfo {author} {\bibfnamefont {M.}~\bibnamefont
  {B{\l}aszak}}\ and\ \bibinfo {author} {\bibfnamefont {Z.}~\bibnamefont
  {Doma{\'n}ski}},\ }\href@noop {} {\bibfield  {journal} {\bibinfo  {journal}
  {Ann. Phys.}\ }\textbf {\bibinfo {volume} {327}},\ \bibinfo {pages} {167}
  (\bibinfo {year} {2012})},\ \Eprint {http://arxiv.org/abs/arXiv:1009.0150v2
  [math-ph]} {arXiv:1009.0150v2 [math-ph]} \BibitemShut {NoStop}%
\bibitem [{\citenamefont {Kontsevich}(2003)}]{Kontsevich:2003}%
  \BibitemOpen
  \bibfield  {author} {\bibinfo {author} {\bibfnamefont {M.}~\bibnamefont
  {Kontsevich}},\ }\href@noop {} {\bibfield  {journal} {\bibinfo  {journal}
  {Lett. Math. Phys.}\ }\textbf {\bibinfo {volume} {66}},\ \bibinfo {pages}
  {157} (\bibinfo {year} {2003})},\ \Eprint
  {http://arxiv.org/abs/arXiv:q-alg/9709040v1} {arXiv:q-alg/9709040v1}
  \BibitemShut {NoStop}%
\bibitem [{\citenamefont {Gutt}\ and\ \citenamefont
  {Rawnsley}(1999)}]{Gutt:1999}%
  \BibitemOpen
  \bibfield  {author} {\bibinfo {author} {\bibfnamefont {S.}~\bibnamefont
  {Gutt}}\ and\ \bibinfo {author} {\bibfnamefont {J.}~\bibnamefont
  {Rawnsley}},\ }\href@noop {} {\bibfield  {journal} {\bibinfo  {journal} {J.
  Geom. Phys.}\ }\textbf {\bibinfo {volume} {29}},\ \bibinfo {pages} {347}
  (\bibinfo {year} {1999})}\BibitemShut {NoStop}%
\bibitem [{\citenamefont {Doma{\'n}ski}\ and\ \citenamefont
  {B{\l}aszak}(2013)}]{Domanski:2013}%
  \BibitemOpen
  \bibfield  {author} {\bibinfo {author} {\bibfnamefont {Z.}~\bibnamefont
  {Doma{\'n}ski}}\ and\ \bibinfo {author} {\bibfnamefont {M.}~\bibnamefont
  {B{\l}aszak}},\ }\href@noop {} {\enquote {\bibinfo {title} {On equivalence of
  star-products in arbitrary canonical coordinates},}\ } (\bibinfo {year}
  {2013}),\ \bibinfo {note} {eprint arXiv:1305.4026 [math-ph]}\BibitemShut
  {NoStop}%
\bibitem [{\citenamefont {B{\l}aszak}\ and\ \citenamefont
  {Doma{\'n}ski}(2013)}]{Blaszak:2013}%
  \BibitemOpen
  \bibfield  {author} {\bibinfo {author} {\bibfnamefont {M.}~\bibnamefont
  {B{\l}aszak}}\ and\ \bibinfo {author} {\bibfnamefont {Z.}~\bibnamefont
  {Doma{\'n}ski}},\ }\href@noop {} {\bibfield  {journal} {\bibinfo  {journal}
  {Ann. Phys.}\ }\textbf {\bibinfo {volume} {331}},\ \bibinfo {pages} {70}
  (\bibinfo {year} {2013})},\ \Eprint {http://arxiv.org/abs/eprint
  arXiv:1208.2835 [math-ph]} {eprint arXiv:1208.2835 [math-ph]} \BibitemShut
  {NoStop}%
\bibitem [{\citenamefont {Benenti}\ \emph
  {et~al.}(2002{\natexlab{a}})\citenamefont {Benenti}, \citenamefont {Chanu},\
  and\ \citenamefont {Rastelli}}]{Benenti:2002a}%
  \BibitemOpen
  \bibfield  {author} {\bibinfo {author} {\bibfnamefont {S.}~\bibnamefont
  {Benenti}}, \bibinfo {author} {\bibfnamefont {C.}~\bibnamefont {Chanu}}, \
  and\ \bibinfo {author} {\bibfnamefont {G.}~\bibnamefont {Rastelli}},\
  }\href@noop {} {\bibfield  {journal} {\bibinfo  {journal} {J. Math. Phys.}\
  }\textbf {\bibinfo {volume} {43}},\ \bibinfo {pages} {5183} (\bibinfo {year}
  {2002}{\natexlab{a}})}\BibitemShut {NoStop}%
\bibitem [{\citenamefont {Benenti}\ \emph
  {et~al.}(2002{\natexlab{b}})\citenamefont {Benenti}, \citenamefont {Chanu},\
  and\ \citenamefont {Rastelli}}]{Benenti:2002b}%
  \BibitemOpen
  \bibfield  {author} {\bibinfo {author} {\bibfnamefont {S.}~\bibnamefont
  {Benenti}}, \bibinfo {author} {\bibfnamefont {C.}~\bibnamefont {Chanu}}, \
  and\ \bibinfo {author} {\bibfnamefont {G.}~\bibnamefont {Rastelli}},\
  }\href@noop {} {\bibfield  {journal} {\bibinfo  {journal} {J. Math. Phys.}\
  }\textbf {\bibinfo {volume} {43}},\ \bibinfo {pages} {5223} (\bibinfo {year}
  {2002}{\natexlab{b}})}\BibitemShut {NoStop}%
\end{thebibliography}
\end{document}